\documentclass[12pt]{article}
\usepackage{amssymb}
\usepackage{bbm}
\usepackage{epsfig}
\usepackage{array}
\usepackage{float}
\usepackage{dsfont}
\usepackage{amstext}
\usepackage[tbtags]{amsmath}
\usepackage{psfrag}
\usepackage{a4}
\usepackage{a4wide}
\usepackage{hyperref}
\usepackage{multicol} 
\usepackage{multirow}
\usepackage{subcaption}
\usepackage{ulem}
\usepackage {setspace}
\usepackage{xcolor,colortbl}
\def\be{\begin{equation}}
\def\ee{\end{equation}}
\def\bea{\begin{eqnarray}}
\def\eea{\end{eqnarray}}
\DeclareMathOperator{\Tr}{Tr}
\DeclareMathOperator{\diag}{diag}
\newcolumntype{g}{>{\columncolor{Gray}}c}
\newcolumntype{w}{>{\columncolor{Gray2}}c}
\definecolor{Gray}{gray}{0.95}
\definecolor{Gray2}{gray}{0.98}
\begin{document}

\begin{center}
\baselineskip 20pt 
{\Large\bf Radiative Symmetry breaking, Cosmic Strings and Observable Gravity Waves in $U(1)_R$ symmetric $SU(5) \times U(1)_{\chi}$}
\vspace{1cm}

{\large 
	\textbf{Waqas Ahmed}$^{a}$ \footnote{E-mail: \texttt{\href{mailto:waqasmit@nankai.edu.cn}{waqasmit@hbpu.edu.cn}}}
	and \textbf{Umer Zubair}$^{b}$ \footnote{E-mail: \texttt{\href{mailto:umer@udel.edu}{umer@udel.edu}}}
} 
\vspace{.5cm}

{\baselineskip 20pt \it
	$^{a}$ \it
	School of Mathematics and Physics, \\ Hubei Polytechnic University, Huangshi 435003, China \\
	\vspace*{6pt}
	$^b$Division of Science and Engineering,  \\ Pennsylvania State University, Abington, PA 19001, USA \\
	\vspace{2mm} }

\vspace{1cm}
\end{center}

\begin{abstract}
We implement shifted hybrid inflation in the framework of supersymmetric $SU(5) \times U(1)_{\chi}$ GUT model which provides a natural solution to the monopole problem appearing in the spontaneous symmetry breaking of $SU(5)$. The $U(1)_{\chi}$ symmetry is radiatevely broken after the end of inflation at an intermediate scale, yielding topologically stable cosmic strings. The Planck's bound on the gravitational interaction strength of these strings, characterized by $G_N \mu_s$ are easily satisfied with the $U(1)_{\chi}$ symmetry breaking scale which depends on the initial boundary conditions at the GUT scale. The dimension-5 proton lifetime for the decay $p \rightarrow K^+ \bar{\nu}$, mediated by color-triplet Higgsinos is found to satisfy current Super-Kamiokande bounds for SUSY breaking scale $M_{\text{SUSY}} \gtrsim 12.5$ TeV. We show that with minimal K\"ahler potential, the soft supersymmetry breaking terms play a vital role in bringing the scalar spectral index $n_s$ within the Planck's latest bounds, although with small tensor modes $r \lesssim 2.5 \times 10^{-6}$ and $SU(5)$ gauge symmetry breaking scale in the range ($2 \times 10^{15} \lesssim M_{\alpha} \lesssim 2 \times 10^{16}$) GeV. By employing non-minimal terms in the K\"ahler potential, the tensor-to-scalar ratio approaches observable values ($r \lesssim 10^{-3}$) with the $SU(5)$ symmetry breaking scale $M_{\alpha} \simeq 2 \times 10^{16}$ GeV.  
\end{abstract}

%
\section{\large{\bf Introduction}}

SUSY hybrid inflation \cite{Dvali:1994ms,Copeland:1994vg,Linde:1993cn,Dvali:1997uq} provides fascinating framework to realize inflation within the grand unified theories (GUTs) of particle physics. Several GUT models such as, $SU(5)$ \cite{Georgi:1974sy}, Flipped $SU(5)$ \cite{Barr:1981qv,Antoniadis:1987dx,Rehman:2018nsn} and the Pati-Salam symmetry $SU(4)_C \times SU(2)_L \times SU(2)_R$ \cite{Pati:1974yy,Mohapatra:1974gc,Ahmed:2018jlv}, have been employed successfully to realize standard, shifted and smooth variants of hybrid inflation \cite{Ahmed:2022vlc,Rehman:2012gd,Rehman:2014rpa,Khalil:2010cp,urRehman:2006hu,Ahmed:2022wed}. The $SU(5) \times U(1)_{\chi}$ gauge symmetry is another suitable choice as a GUT model due to its various attractive features \cite{Kibble:1982ae,apal:2019}. The whole gauge group of the model is embedded in $SU(5) \times U(1)_{\chi} \subset SO(10)$, owing to special $U(1)_{\chi}$ charge assignment  \cite{Kibble:1982ae}. In contrast to the $SU(5)$ \cite{Georgi:1974sy} model, a discrete $Z_2$ symmetry that avoids rapid proton decay, naturally arises after the breaking of $U(1)_{\chi}$ factor. This $Z_2$ symmetry not only serves as the Minimal Supersymmetric Standard Model (MSSM) matter parity, but also ensures the existence of a stable lightest supersymmetric particle (LSP) which can be a viable cold dark matter candidate. Furthermore, the right-handed neutrino mass is naturally generated by the breaking of $U(1)_{\chi}$ symmetry after one of the fields carrying $U(1)_{\chi}$ charge acquires a Vacuum Expectation Value (VEV) at some intermediate scale. The well-known advantages of $U(1)_{\chi}$ symmetry include seesaw physics to explain neutrino oscillations, and baryogenesis via leptogenesis \cite{Fukugita:1986hr,Lazarides:1990huy}.

The breaking of $SU(5)$ part of the gauge symmetry leads to copious production of magnetic monopoles \cite{Hill:1982iq} in conflict with the cosmological observations whereas, the breaking of $U(1)_{\chi}$ factor yields topologically stable cosmic strings \cite{Kibble:1982ae,Vilenkin:1984ib,Vilenkin:2000jqa}. The cosmic strings can be made to survive if $U(1)_{\chi}$ breaks after the end of inflation. In order to avoid the undesired monopoles, the shifted or smooth variant of hybrid inflation \cite{Khalil:2010cp} can be employed, where the gauge symmetry is broken during inflation and disastrous monopoles are inflated away. In the simplest SUSY hybrid inflationary scenario the potential along the inflationary track is completely flat at tree level. The inclusion of radiative corrections (RC) to the scalar potential provide necessary slope needed to drive inflaton towards the SUSY vacuum and in doing so the gauge symmetry $G$ breaks spontaneously to its subgroup $H$.

In this paper, we implement shifted hybrid inflation scenario in the $SU(5) \times U(1)_{\chi}$ GUT model \cite{apal:2019} where the $SU(5)$ symmetry is broken during inflation and the $U(1)_{\chi}$ symmetry radiatevely breaks to its $Z_2$ subgroup at some intermediate scale. The scalar spectral index $n_s$ lies in the observed range of Planck's results \cite{Planck:2018jri} provided the inflationary potential incorporates either the soft supersymmetry (SUSY) breaking terms~\cite{rehman,gravitywaves,Ahmed:2022rwy,Afzal:2022vjx,Ahmed:2021dvo,Ahmed:2022rwy}, or higher-order terms in the K\"ahlar potential \cite{urRehman:2006hu,bastero}. Without these terms, the scalar spectral index $n_s$ lies close to 0.98 which is acceptable only if the effective number of light neutrino species are slightly greater than 3 \cite{Ade:2015lrj}. We show that, by taking soft SUSY contribution into account along with the supergravity (SUGRA) corrections in a minimal K\"ahlar potential setup, the predictions of the model are consistent with the Planck’s latest bounds on scalar spectral index $n_s$ \cite{Ade:2015lrj}, although the values of tensor to scalar ratio remain small. By employing non-minimal K\"ahler potential, large tensor modes are easily obtained, approaching observable values potentially measurable by near-future experiments such as, PRISM \cite{Andre:2013afa}, LiteBird \cite{Matsumura:2013aja}, CORE \cite{Finelli:2016cyd}, PIXIE \cite{Kogut:2011xw}, CMB-S4 \cite{Abazajian:2019eic}, CMB-HD \cite{Sehgal:2019ewc} and PICO \cite{SimonsObservatory:2018koc}. Moreover, the $U(1)_{\chi}$ symmetry radiatively breaks after the end of inflation at an intermediate scale, yielding topologically stable cosmic strings. The Planck's bound \cite{Ade:2013xla,Ade:2015xua} on the strength of gravitational interaction of the strings, $G_N \mu_s$ are easily satisfied with the $U(1)_{\chi}$ symmetry breaking scale obtained in the model, which depends on the initial boundary conditions at the GUT scale. Furthermore, the Super-Kamiokande bounds \cite{Super-Kamiokande:2016exg} on dimension-5 proton decay lifetime are easily satisfied for SUSY breaking scale $M_{\text{SUSY}} \gtrsim 12.5$ TeV.

The rest of the paper is organised as follows. Sec. \ref{sec2} provides the description of the $SU(5) \times U(1)_{\chi}$ model. The implementation of shifted hybrid inflation including the mass spectrum of the model, gauge coupling unification and dimension-5 proton decay is discussed in Sec. \ref{sec3}. The results and inflationary predictions of the model with minimal K\"ahler potential are presented in Sec. \ref{sec5} and with non-minimal K\"ahler potential in Sec. \ref{sec6}. The radiative breaking of $U(1)_{\chi}$ symmetry and cosmic strings is discussed in Sec. \ref{sec7}. Finally we summarize our results in Sec. \ref{sec8}.

%
\section{\large{The $U(1)_R$ Symmetric $SU(5) \times U(1)_{\chi}$ Model}} \label{sec2}%
The $\mathbf{10}$, $\bar{\mathbf{5}}$ and $\mathbf{1}$ dimensional representations of the group $SU(5)$ constitute the $\mathbf{16}$ (spinorial) representation of $SO(10)$ and contains the MSSM matter superfields. Their decomposition with respect to the MSSM gauge symmetry is 
\begin{equation}
	\label{Mspectrum} 
	\begin{split}
F_i &\equiv (\mathbf{10}, -1) = Q(\mathbf{3}, \mathbf{2}, 1/6) + u^c(\bar{\mathbf{3}}, \mathbf{1}, -2/3) + e^c (\mathbf{1}, \mathbf{1}, 1)~,   \\
\bar{f}_i &\equiv (\bar{\mathbf{5}}, +3) = d^c (\bar{\mathbf{3}}, \mathbf{1}, 1/3) + \ell(\mathbf{1}, \mathbf{2}, -1/2)~,    \\
\nu_{i}^{c} &\equiv (\mathbf{1}, -5) =  \nu^c(\mathbf{1}, \mathbf{1}, 0)~,
	\end{split}
\end{equation}
where $i = 1, 2, 3$ denotes the generation index. The scalar sector of $SU(5) \times U(1)_{\chi}$ consists of the following superfields: a pair of Higgs fiveplets, $h\,\equiv (\mathbf{5}, 2)$, $\bar{h}\,\equiv (\bar{\mathbf{5}}, -2)$, containing the electroweak Higgs doublets ($h_d, h_u$) and color Higgs triplets ($D_h,\bar{D}_{\bar{h}}$); a Higgs superfield $\Phi$ that belongs to the adjoint representation ($\Phi\, \equiv 24_{0}$) and responsible for breaking $SU(5)$ gauge symmetry to MSSM gauge group; a pair of superfields ($\chi$, $\bar{\chi}$) which trigger the breaking of $U(1)_{\chi}$ into a $Z_2$ symmetry which is realized as the MSSM matter parity; and finally, a gauge singlet superfield $S$ whose scalar component acts as an inflaton.
The decomposition of the above $SU(5)$ representations under the MSSM gauge group is
\begin{align}
	\label{Hspectrum} 
	\begin{split}
\Phi  \equiv {}& (\mathbf{24}, 0) = \Phi_{24}(\mathbf{1}, \mathbf{1}, 0) + W_H (\mathbf{1}, \mathbf{3}, 0) + G_H (\mathbf{8}, \mathbf{1}, 0)  \\
& \qquad  \quad+  Q_H(\mathbf{3}, \mathbf{2}, -5/6) + \bar{Q}_H(\mathbf{3}, \mathbf{2}, 5/6),   \\ 
h  \equiv {}& (\mathbf{5}, 2) = D_h(\mathbf{3}, \mathbf{1}, -1/3) + h_u (\mathbf{1}, \mathbf{2}, 1/2)~,   \\
\bar{h}  \equiv {}& (\bar{\mathbf{5}}, -2) = \bar{D}_{\bar{h}}(\bar{\mathbf{3}}, \mathbf{1}, 1/3) + h_d(\mathbf{1}, \mathbf{2}, -1/2), \\
\chi \equiv {}& (\mathbf{1}, 10), \quad \bar{\chi} \equiv (\mathbf{1}, -10), \quad S \equiv (\mathbf{1}, 0),
	\end{split}
\end{align}
where the singlets ($\chi$, $\bar{\chi}$) originate from the decomposition of $\mathbf{126}$ representation of $SO(10)$
\begin{equation}
	\mathbf{126} = (\mathbf{1}, -10) + (\bar{\mathbf{5}}, -2) + (\mathbf{10}, -6) + (\bar{\mathbf{15}}, 6) + (\mathbf{45}, 2) + (\bar{\mathbf{50}}, -2).
\end{equation}
\begin{table}[t]
	\setlength\extrarowheight{5pt}
	\centering
	\begin{tabular}{c c c}
		\hline \hline \rowcolor{Gray}
		\multicolumn{1}{c}{}                              & \multicolumn{1}{c}{}                                                                                        & \multicolumn{1}{c}{}                                                                        \\ \rowcolor{Gray}
		\multicolumn{1}{c}{\multirow{-2}{*}{Superfields}} &  \multicolumn{1}{c}{\multirow{-2}{*}{\begin{tabular}[c]{@{}c@{}}Representations under\\ $SU(5) \times U(1)_{\chi}$\end{tabular}}} & \multicolumn{1}{c}{\multirow{-2}{*}{\begin{tabular}[c]{@{}c@{}}Global\\ $U(1)_R $\end{tabular}}} \\  \hline
	\rowcolor{Gray2} 	\multicolumn{3}{c}{Matter sector}                                      \\ \hline
		\multicolumn{1}{c}{$F_i$}     &       $\left( \mathbf{10}, -1 \right)$     &    $3/10$ \\
		\multicolumn{1}{c}{$\bar{f}_i$} &  $\left( \bar{\mathbf{5}}, 3 \right)$ &  $1/10$\\
		\multicolumn{1}{c}{$\nu_{i}^c$}	&   $\left( \mathbf{1}, -5 \right)$ & $1/2$\\ \hline 
	\rowcolor{Gray2} 	\multicolumn{3}{c}{Scalar sector}                               \\ \hline
		\multicolumn{1}{c}{$\Phi$}	&   $\left( \mathbf{24}, 0 \right)$ &  $0$\\
		\multicolumn{1}{c}{$h$}	&   $\left( \mathbf{5}, 2 \right)$ & $2/5$\\
		\multicolumn{1}{c}{$\bar{h}$} & $\left( \bar{\mathbf{5}}, -2 \right)$ & $3/5$\\
		\multicolumn{1}{c}{$\chi$}	&   $\left( \mathbf{1}, 10 \right)$ & $0$\\
		\multicolumn{1}{c}{$\bar{\chi}$} & $\left( \mathbf{1}, -10 \right)$ & $0$\\
		\multicolumn{1}{c}{$S$}	& $\left( \mathbf{1}, 0 \right)$ & $1$\\  \hline \hline
	\end{tabular}
	\caption{The representations of matter and scalar superfields under $SU(5) \times U(1)_{\chi}$ gauge symmetry and global $U(1)_R$ symmetry in shifted hybrid inflation model.}
	\label{tab:field_charges}
\end{table}
Following \cite{Khalil:2010cp}, the $U(1)_R$ charge assignment of the superfields is given in Table \ref{tab:field_charges} along with their transformation properties. 
The  $SU(5) \times U(1)_{\chi}$ and $U(1)_R$, symmetric superpotential of the model with the leading-order non-renormalizable terms is given by 
\bea%
W &=& S\left[\kappa M^2-\kappa \Tr(\Phi^{2})-\frac{\beta}{m_P}
\Tr(\Phi^3) + \sigma_{\chi} \chi \bar{\chi}\right] + \gamma \bar{h} \Phi h + \delta \bar{h} h  \nonumber \\
&+& y_{ij}^{(u)}\,F_i\,F_j\,h + y_{ij}^{(d,e)}\,F_i\,\bar{f}_j\,\bar{h}
+y_{ij}^{(\nu)}\,\nu_{i}^c\,\bar{f}_j\,h + \lambda_{ij} \chi \nu_{i}^{c} \nu_{j}^{c} , \label{superpotential}
\eea %
where $M$ is a superheavy mass and $m_P = 2.43 \times 10^{18}$ GeV is the reduced Planck mass. The terms in square bracket in the first line are relevant for shifted hybrid inflation while, the last two terms are involved in the solution of doublet-triplet splitting problem, as discussed in section \ref{sec4}.
The Yukawa couplings $y_{ij}^{(u)}$, $ y_{ij}^{(d,e)}$, $y_{ij}^{(\nu)}$ in the second line of \eqref{superpotential} generate Dirac masses for quarks and leptons after the electroweak symmetry breaking, whereas $m_{\nu_{ij}} = \lambda_{ij} \langle \chi \rangle$ is the right-handed neutrino mass matrix, generated after $\chi$ acquires a VEV through radiative breaking of $U(1)_{\chi}$ symmetry, as discussed in Sec. \ref{sec7}.  

The superpotential $W$ exhibits a number of interesting features as a consequence of global $U(1)_{R}$ symmetry. First, it allows only linear terms in $S$ in the superpotential, omitting the higher order ones, such as $S^2$ which could generate an inflaton mass of Hubble size, invalidating the inflationary scenario. Second, the $U(1)_{R}$ symmetry naturally avoids the so called $\eta$ problem \cite{Linde:1997sj}, that appears when SUGRA corrections are included. Finally, several dangerous dimension-5 proton decay operators are highly suppressed.

\section{\large{\bf Shifted Hybrid $SU(5) \times U(1)_{\chi}$ Inflation}}\label{sec3}
In this section, the effective scalar potential is computed considering contributions from the $F$- and $D$-term sectors. The superpotential terms relevant for shifted hybrid inflation are
\bea%
W &\supset& S\left[\kappa M^2-\kappa \Tr(\Phi^{2})-\frac{\beta}{m_P}
\Tr(\Phi^3)\right]  + \gamma \bar{h} \Phi h + \delta \bar{h} h \nonumber \\
&+&  \sigma_{\chi} S  \chi \bar{\chi} +\lambda_{ij} {\chi} \nu_{i}^{c} \nu_{j}^{c}~. \label{superpotential_inflation}
\eea %
In component form, the above superpotential is expanded as follows,%
\bea%
W \supset S\left[\kappa M^2-\kappa\frac{1}{2}
\sum_{i}\phi_{i}^{2}-\frac{\beta}{4 m_P}d_{ijk}\phi_{i}\phi_{j}
\phi_{k}\right] &+& \delta \bar{h}_{a}h_{a} + \gamma T_{a b}^{i}\phi^{i}\bar{h}_{a} h_{b} \nonumber \\
&+& \sigma_{\chi} S  \chi \bar{\chi} +\lambda_{ij} {\chi} \nu_{i}^{c}\nu_{j}^{c},
\label{superpot-shift}%
\eea %
where $\Phi = \phi_i T^i$ with Tr$[T_i T_j] = \frac{1}{2}\delta_{ij}$ and $d_{ijk} = 2$Tr$[T_i\{T_j,T_k\}]$ in the $SU(5)$ adjoint basis. The $F$-term scalar potential obtained from the above superpotential is given by %
\begin{eqnarray}%
V_F &=&  \left| \; \kappa M^2-\kappa\frac{1}{2}
\sum_{i}\phi_{i}^{2}-\frac{\beta}{4 m_P}d_{ijk}\phi_{i}\phi_{j}
\phi_{k}+  \sigma_{\chi}  \chi \bar{\chi}   \; \right|^{2}  \nonumber \\ 
&+& \sum_{i}\left|\kappa S \phi_{i}+\frac{3
	\beta}{4 m_P}d_{ijk} S \phi_{j} \phi_{k}
- \gamma T_{a
	b}^{i}\bar{h_{a}}h_{b}\right|^{2} \nonumber \\ 
&+& \sum_{b}\left|\gamma
T_{a b}^{i}\phi^{i}\bar{h_{a}}+\delta
\bar{h_{b}}\right|^{2}+\sum_{b}\left|\gamma T_{a
	b}^{i}\phi^{i}h_{a}+\delta
h_{b}\right|^{2} \nonumber\\
&+&  \left| \, \sigma_{\chi} S \bar{\chi} + \lambda_{ij}  \nu_{i}^{c}\nu_{j}^{c}\, \right|^2 + \left| \, \sigma_{\chi} S \chi \, \right|^2 + \left|2 \lambda_{ij} {\chi} \nu_{i}^{c}\right|^2 ,
\label{scalarpot-shift}
\end{eqnarray}%
where the scalar components of the superfields are denoted by the same symbols as the corresponding superfields. The VEV's of the fields at the global SUSY minimum of the above potential are given by,

\begin{gather}
S^0 = h_{a}^0  =  \bar{h_{a}^0} = \nu_{i}^{c\,0}=0,  \;\; \chi^0 = \bar{\chi}^0 = 0
\label{gmin}
\end{gather} 
with $\phi_i^0$ satisfying the following equation: %
\be%
\sum_{i=1}^{24}(\phi_{i}^{0})^{2} + \frac{\beta}{2 \kappa m_P}
d_{ijk}
\phi^0_i \phi^0_j \phi^0_k =2M^{2}.
\ee%
The superscript `0' denotes the field value at its global minimum. The superfield pair ($\chi, \bar{\chi}$) break $U(1)_{\chi}$ to $Z_2$, the matter parity. This symmetry ensures the existence of a  lightest supersymmetric particle (LSP) which could play the role of cold dark matter. Further, as discussed in~\cite{apal:2019}, this $Z_2$  symmetry yields topologically stable cosmic strings. 

Using $SU(5)$ symmetry transformation the VEV matrix $\Phi^0 = \phi_i^0 T^i$ can be aligned in the $24$-direction,
\be
\Phi_{24}^0 = \frac{\phi_{24}^0}{\sqrt{15}} \left( 1, 1, 1, - 3/2, - 3/2 \right).
\ee
Thus the $SU(5)$ gauge symmetry is broken down to Standard Model gauge group $G_{\text{SM}}$ by the non-vanishing VEV of $\phi_{24}^0$ which is a singlet under $G_{\text{SM}}$.

The $D$-term scalar potential,
\bea %
V_D &=& \frac{g_{5}^2}{2} \sum_{i} \left( f^{ijk} \phi_j \phi_{k}^{\dagger}  + T^i \left( \left| h_a \right|^2  - \left| \bar{h}_a \right|^2  \right) \right)^2 \nonumber \\ 
&+& \frac{g_{\chi}^2}{2} \left( q_{\chi} \left| \chi \right|^2 + q_{\bar{\chi}} \left| \bar{\chi} \right|^2 + \left( q_{\bar{\chi}} + q_{\chi} \right) \varsigma \right)^2,
\label{dterm}
\eea %
also vanishes for this choice of the VEV (since $f^{i, 24, 24} = 0$) and for $\vert \bar{h}_a \vert = \vert h_a \vert$, $\vert \bar{\chi} \vert = \vert \chi \vert$. 

The scalar potential in Eq.~(\ref{scalarpot-shift}) can be written in terms of the dimensionless variables%
\be %
z = \frac{|S|}{M}~, ~~~~~~~~~~~~ 
y = \frac{\phi_{24}}{M\sqrt{2}},~~~~~~~~~~~~ %
\ee %
as follows, 
\be %
\tilde{V} = \frac{V}{\kappa^2 M^4}  = \left( 1-y^2+\alpha y^3\right)^2 + 2 z^2 y^2\left(1-3\alpha y/2\right)^2   ~,%
\label{VF0}
\ee %
\begin{figure}[t]
	\centering 
	\begin{subfigure}[b]{0.495\textwidth}
		\centering
		\caption{Standard, $\alpha = 0$}
		\includegraphics[width=8cm]{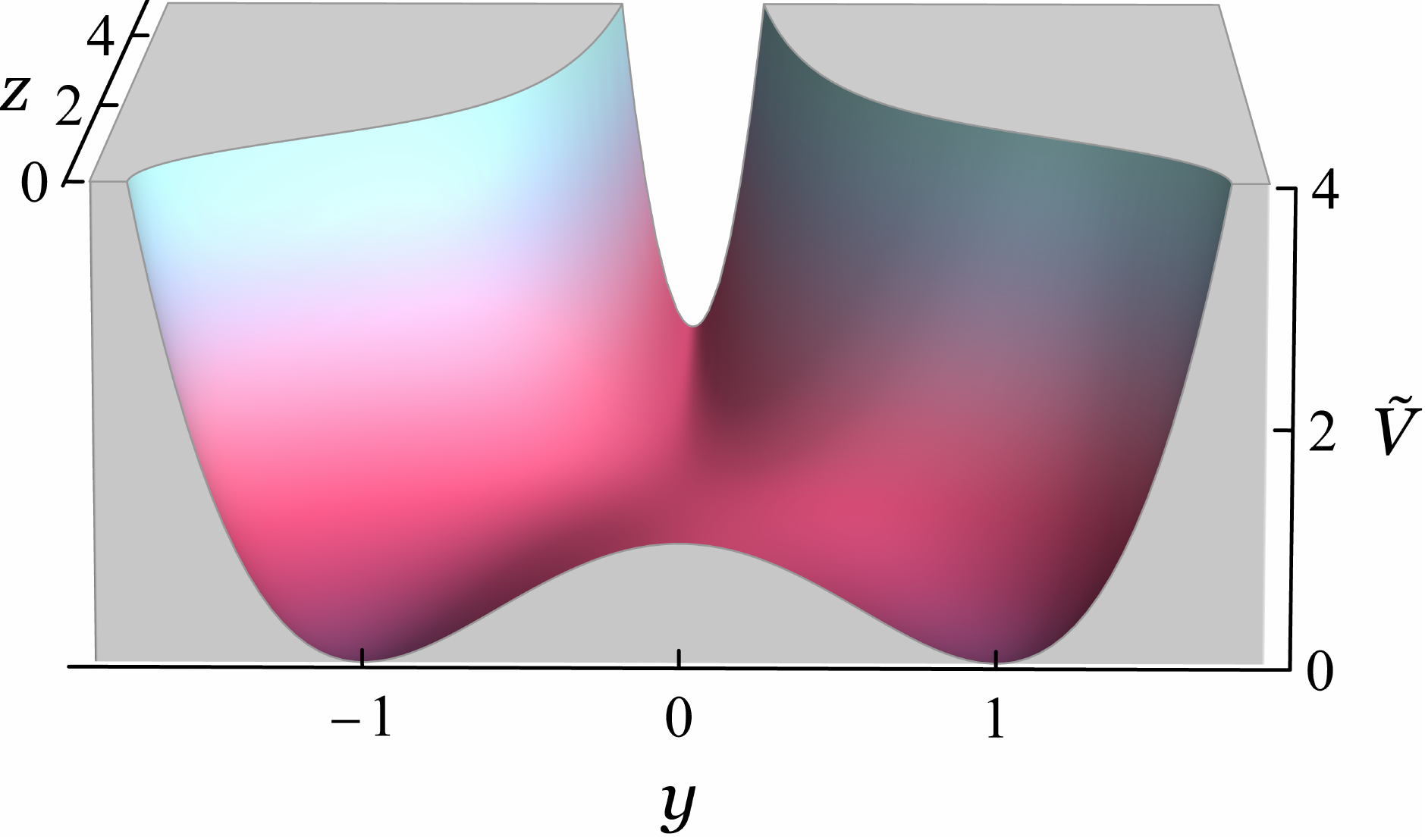}
		\label{fig:alpha_0}
	\end{subfigure}
	\begin{subfigure}[b]{0.495\textwidth}
		\caption{Shifted, $\alpha = 0.25$}
		\centering \includegraphics[width=8cm]{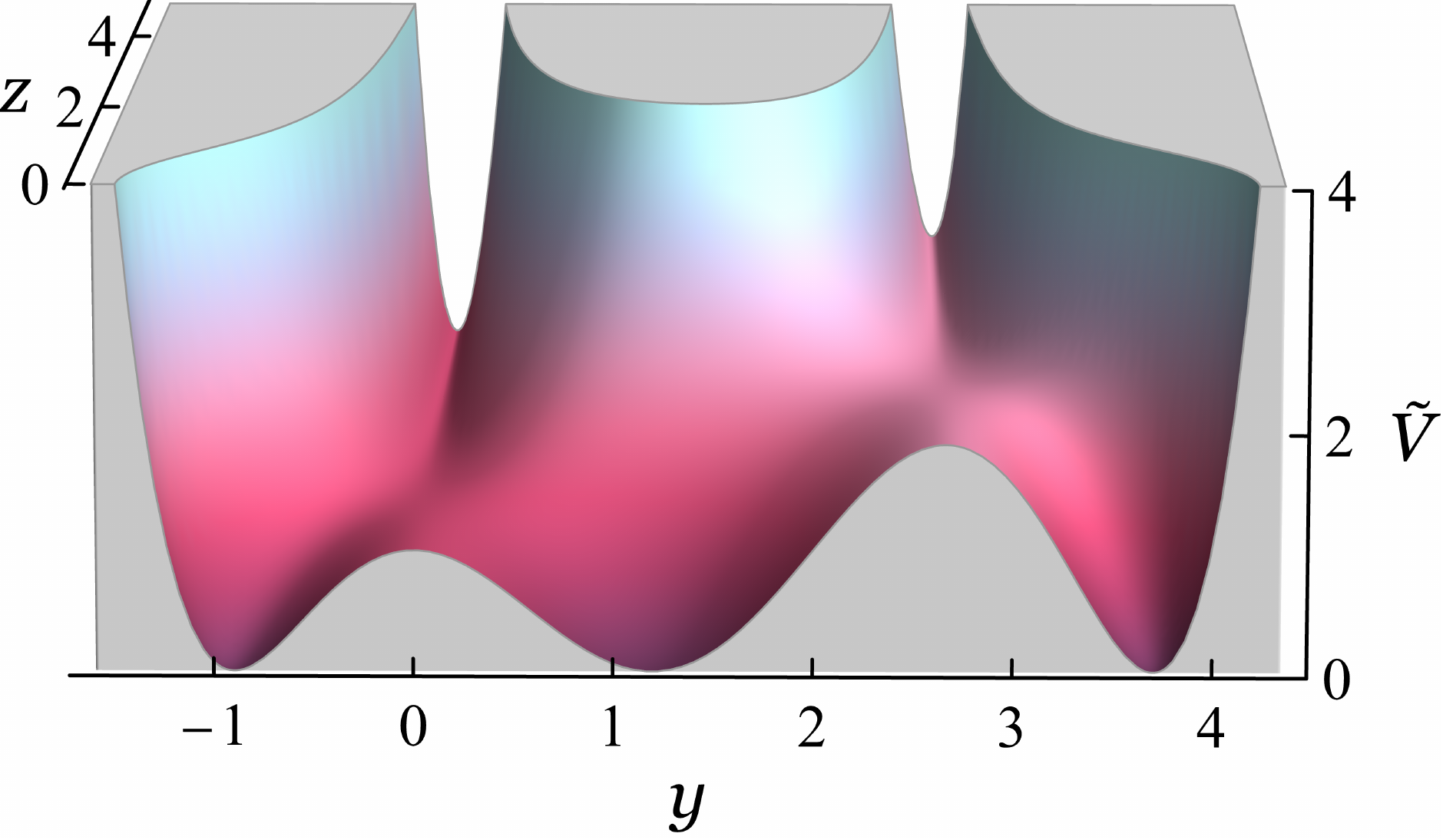}
		\label{fig:alpha_025}
	\end{subfigure}
	\begin{subfigure}[b]{0.495\textwidth}
		\caption{Shifted, $\alpha = 0.3$}
		\centering \includegraphics[width=8cm]{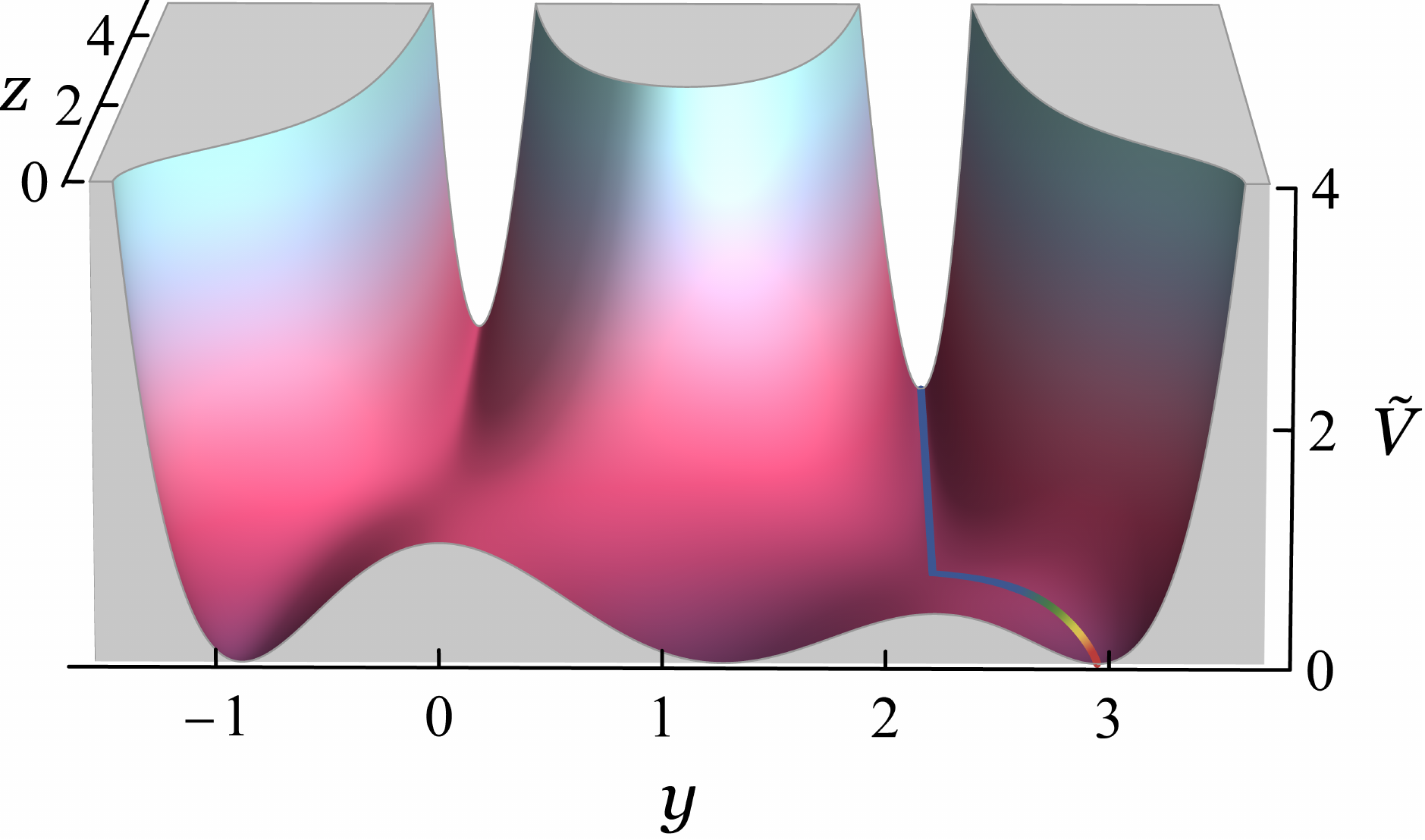}
		\label{fig:alpha_03}
	\end{subfigure}
	\caption{The tree-level, global dimensionless scalar potential $\tilde{V}=V/\kappa^2M^4$ versus $y$ and $z$ for; $\alpha = 0$ (a), $\alpha = 0.25$ (b) and $\alpha = 0.3$ (c). The standard track corresponds to $\alpha = 0$ whereas, $\alpha \neq 0$ corresponds to two shifted trajectories. The inflationary track feasible for realizing successful inflation is shown in panel (b) for $\alpha = 0.3$.}
	\label{fig1}
\end{figure}
where $\alpha = \beta M/ \sqrt{30}\, \kappa m_P$. This dimensionless potential exhibits the following three extrema
\begin{equation}
y_1=0, \label{1st minima}
\end{equation}
\be
y_2=\frac{2}{3 \alpha }, \label{2nd minima}
\ee
\begin{eqnarray}
y_3 &=& \frac{1}{3\alpha} + \frac{1}{3 \sqrt[3]{2} \alpha }\Bigg( \sqrt[3]{2 -27 \alpha ^2+\sqrt{\left(2-27 \alpha ^2\right)^2+4 \left(9 \alpha ^2 z^2-1\right)^3}}   \nonumber\\ 
&-&    \sqrt[3]{-2 + 27 \alpha ^2+\sqrt{\left(2-27 \alpha ^2\right)^2+4 \left(9 \alpha ^2 z^2-1\right)^3}} \Bigg).
\end{eqnarray}
for a constant value of $z$ and is displayed in Fig. \ref{fig1} for different values of $\alpha$. The first extremum $y_1$ with $\alpha = 0$ corresponds to the standard hybrid inflation for which $y = 0, z > 1$ is the only inflationary trajectory that evolves at $z = 0$ into the global SUSY minimum at $y = \pm 1$ (Fig. \ref{fig:alpha_0}). For $\alpha \neq 0$, a shifted trajectory appears at $y = y_2$, in addition to the standard trajectory at $y = y_1 =0$, which is a local maximum (minimum) for $z < \sqrt{4/27\alpha^2-1}$ ($z > \sqrt{4/27\alpha^2-1}$). For $\alpha < \sqrt{2/27}\simeq 0.27$, this shifted trajectory lies higher than the standard trajectory (Fig. \ref{fig:alpha_025}). In order to have suitable initial conditions for realizing inflation along the shifted track, we assume $\alpha > \sqrt{2/27}$, for which the shifted trajectory lies lower than the standard trajectory (Fig. \ref{fig:alpha_03}). Moreover, to ensure that the shifted inflationary trajectory at $y_2$ can be realized before $z$ reaches zero, we require $\alpha < \sqrt{4/27} \simeq 0.38$. Thus, for $0.27 < \alpha < 0.38$, while the inflationary dynamics along the shifted track remain the same as for the standard track, the $SU(5)$ gauge symmetry is broken during inflation, hence alleviating the magnetic monopole problem. As the inflaton slowly rolls down the inflationary valley and enters waterfall regime at $z =\sqrt{4/27\alpha^2-1}$, its fast rolling ends inflation, and the system starts oscillating about the vacuum at $z= 0$ and $y=y_3$. In order to calculate one-loop radiative correction along $y_2$, we need to compute the mass spectrum of the model along this track where both $SU(5)$ gauge symmetry and SUSY are broken.

During inflation, the field $\Phi$ acquires a VEV in the $\phi_{24}$ direction which breaks the $SU(5)$ gauge symmetry down to SM gauge group $G_{SM}$ while, the $U(1)_{\chi}$ symmetry remains unbroken. The potential in Eq. \eqref{scalarpot-shift} generates the following masses for: 2 real scalars 
\begin{equation}
	m_{24_\pm}^2 = \pm \kappa^2 M_{\alpha}^2 + \kappa^2 \vert S \vert^2 ,
\end{equation}
22 real scalars
\begin{table}[t]
	\setlength\extrarowheight{2pt}
	\addtolength{\tabcolsep}{7pt} 
	\centering
	\begin{tabular}{lr}
		\hline\hline \rowcolor{Gray} {\bf Fields} &  {\bf Squared Masses} \\
		\hline  2 Majorana fermions & $\sigma_{\chi}^2 \vert S\vert^2$ \\
		4 real and pseudo scalars & $\sigma_{\chi}^2 \vert S\vert^2 \pm  \kappa \sigma_{\chi} M_{\alpha}^2$ \\
		2 real scalars & $\kappa^2 \vert S\vert^2 \pm \kappa^2 M_{\alpha}^2$ \\
		1 Majorana fermion & $\kappa^2 \vert S\vert^2 $ \\
		22 real scalars & $25 \kappa^2 \vert S\vert^2 \pm 5\kappa^2 M_{\alpha}^2$ \\
		11 Majorana fermions & $25 \kappa^2 \vert S\vert^2 $ \\
		12 real scalars & $ \frac{25}{30} g_5^2 v_2^2 $ \\
		12 Dirac fermions & $\frac{25}{30} g_5^2 v_2^2 $ \\
		12 gauge bosons & $\frac{25}{30} g_5^2 v_2^2 $  \\ 
		\hline
	\end{tabular}%
	\caption{The mass spectrum of the shifted hybrid $SU(5) \times U(1)_{\chi}$ model along the inflationary trajectory $y_2$.} \label{mass_spectrum_tab}
\end{table}
\begin{equation}
	m_{i_\pm}^2 = \pm 5 \kappa^2 M_{\alpha}^2 + 25 \kappa^2 \vert S \vert^2 , \quad i = 1, \dots , 8, 21, 22, 23 ,
\end{equation}
and 4 real and pseudo scalars
\begin{equation}
	m_{(\chi, \bar{\chi})_\pm}^2 = \pm  \kappa \sigma_{\chi} M_{\alpha}^2 +  \sigma_{\chi}^2 \vert S \vert^2, 
\end{equation}
where $M_{\alpha}^2 = M^2 \left( \frac{4}{27 \alpha^2}  - 1\right)$. The superpotential \eqref{superpot-shift} generates: a Majorana fermion with mass squared,
\begin{equation}
	m_{24}^2 =  \kappa^2 \vert S \vert^2 ,
\end{equation}
11 Majorana fermions with mass squared, 
\begin{equation}
	m_{i}^2 = 25 \kappa^2 \vert S \vert^2 , \quad i = 1, \dots , 8, 21, 22, 23,
\end{equation}
and 2 Majorana fermions with mass squared,
\begin{equation}
	m_{\chi, \bar{\chi}}^2 =  \sigma_{\chi}^2 \vert S \vert^2 .
\end{equation}
The scalar fields $\phi_i$ ($i = 9, ...., 20$) obtain a universal mass squared $\frac{25}{30} g_5^2 \upsilon_2^2$, from the $D$-term in Eq. \eqref{dterm}, while the mixing between chiral fermions $\psi_i$ ($i = 9, ...., 20$) and gauginos $\lambda_i$ ($i = 9, ...., 20$) yields 12 Dirac fermions with a mass squared $\frac{25}{30} g_5^2 \upsilon_2^2$. Finally the 12 guage bosons $A_{\mu}^i$ ($i = 9, ...., 20$) obtain a universal mass squared $\frac{25}{30} g_5^2 \upsilon_2^2$. The mass spectrum of the above model, along the shifted inflationary track, is summarized in Table \ref{mass_spectrum_tab}.

The 1-loop radiative correction to the inflationary effective potential is given by %
\bea %
V_{\rm 1 loop} \!\!&\!=\!&\!\! \kappa^2 M_{\alpha}^4 \left(
\frac{\kappa^2}{16 \pi^2} \left[ F(M_{\alpha}^2,x^2)
+11\times 25\,F(5 M_{\alpha}^2,5\,x^2) \right] + \frac{\sigma_{\chi}^2}{8 \pi^2} F(M_{\alpha}^2,y^2) \right),%
\label{Vloop}
\eea%
with
\be %
F(M_{\alpha}^2,x^2) = 
\frac{1}{4}\left( \left( x^4+1\right) \ln \frac{\left( x^4-1\right) 
}{x^4}+2x^2\ln \frac{x^2+1}{x^2-1}+2\ln \frac{\kappa ^{2}M_{\alpha}^{2}x^2}{%
	Q^{2}}-3\right),
\ee%
\be %
F(M_{\alpha}^2,y^2) = 
\frac{1}{4}\left( \left( y^4 + 1\right) \ln \frac{\left( y^4 - 1\right) 
}{y^4}+2y^2\ln \frac{y^2 + 1}{y^2 - 1} + 2\ln \frac{\kappa \sigma_{\chi} M_{\alpha}^{2} y^2}{%
	Q^{2}}-3\right),
\ee%
where $x=|S|/ M_{\alpha}$, $y = \zeta x$, $\zeta = \kappa / \sigma_{\chi}$ and $Q$ is the renormalization scale.

Considering gravity-mediated SUSY breaking scenario, where SUSY is broken in the hidden sector and is communicated gravitationally to the observable sector, the soft potential is \cite{Nilles:1983ge}
\bea
V_{\text{Soft}} = M_{z_{i}}^2\vert z_{i} \vert^2+m_{3/2}\left\{z_{i}W_{i}+\left(A-3\right)W +h.c\right\},
\eea
where $z_{i}$ is observable sector field, $W_{i}=\frac{\partial W}{\partial z_{i}}$, $m_{3/2}$ is the gravitino mass and $A$ is the complex coefficient of the trilinear soft-
SUSY-breaking terms. The effective contributions of soft SUSY breaking terms during inflation can be written as,
\begin{eqnarray}
	V_{\text{Soft}} =  a m_{3/2} \kappa M_{\alpha}^3 x + M_S^2 M_{\alpha}^2  x^2 + \frac{8 M_{\phi}^2 M_{\alpha}^2}{9 \left( 4/27 - \alpha^2 \right)},
\end{eqnarray}
with
\begin{eqnarray}
	a=2\vert A-2 \vert \cos \left(\arg S + \arg \vert A-2 \vert\right) ,
\end{eqnarray}
where $a$ and $M_S$ are the coefficients of soft SUSY breaking linear and mass terms for $S$, respectively, $M_{\phi}$ is the soft mass term for $\phi_{24}$ and $m_{3/2}$ is the gravitino mass.

\subsection{Gauge Coupling Unification} \label{sec2_2}

After the breaking of the $SU(5)$ symmetry, the octet $G_H$ and triplet $W_H$ from the adjoint Higgs field remain massless, as shown in \cite{Barr:2005xya,Fallbacher:2011xg}. The presence of these flat directions is a generic feature of simple groups like the $SU(5)$ with a $U(1)_R$ symmetry, as discussed in \cite{Barr:2005xya,Fallbacher:2011xg}. These fields, however, acquire relatively light masses $\mathcal{O} (\sim \text{TeV})$ from the soft SUSY-breaking terms in our model, which spoils the unification of the gauge couplings. In order to preserve gauge-coupling unification we add the following combination of vectorlike particles 
\begin{equation}
	5 + \bar{5} + E + \bar{E} = \left( D + \bar{D}, L + \bar{L} \right) + E + \bar{E} \,,
\end{equation}
with the $R$-charge, $R \left(5\,\bar{5},E\,\bar{E}\right) = (1,1)$ and allow mass-splitting within a multiplet with some fine tuning. The superpotential of these vectorlike fermions is given by \cite{Masoud:2019gxx},
\begin{eqnarray}
	W &\supset& \frac{\mathcal{A}_{ij}^{(E,\overline{E})}}{m_{P}} \Tr(\Phi^{2}) E_i\overline{E}_j +\frac{\mathcal{B}_{ij}^{(E,\overline{E})}}{m_{P}} \Tr(E_i \Phi^{2}  \overline{E}_j) \nonumber \\
	&+& \frac{\mathcal{A}_{ij}^{(5,\overline{5})}}{m_{P}} \Tr(\Phi^{2}) \Tr(5_i\overline{5}_j) +\frac{\mathcal{B}_{ij}^{(5,\overline{5})}}{m_{P}} \Tr(5_i \Phi^{2}  \overline{5}_j), \nonumber \\
	&\supset&  M_{E} E \overline{E} + M_{D} D \overline{D} + M_{L} L \overline{L}\,.
\end{eqnarray}

Assuming $\mathcal{A}_{ij} = \delta_{ij}\mathcal{A}$ and $\mathcal{B}_{ij}= \delta_{ij}\mathcal{B}$ for convenience, we obtain the following masses of the MSSM field components of vectorlike particles,
\begin{align}
	M_{E}& = \frac{40 \mathcal{A}^{(E,\overline{E})} + 12 \mathcal{B}^{(E,\overline{E})}}{45 \alpha^2}\left( \frac{M^{2}}{m_{P}}\right)\,,\\
	M_{D}& = \frac{60\mathcal{A}^{(5,\overline{5})} + 8 \mathcal{B}^{(5,\overline{5})}}{135 \alpha^2}\left( \frac{M^{2}}{m_{P}}\right)\,,\\
	M_{L}& = \frac{20 \mathcal{A}^{(5,\overline{5})} + 6 \mathcal{B}^{(5,\overline{5})}}{45 \alpha^2}\left( \frac{M^{2}}{m_{P}}\right)\,.
\end{align}
\begin{figure}[t]
	\centering \includegraphics[width=8.00cm]{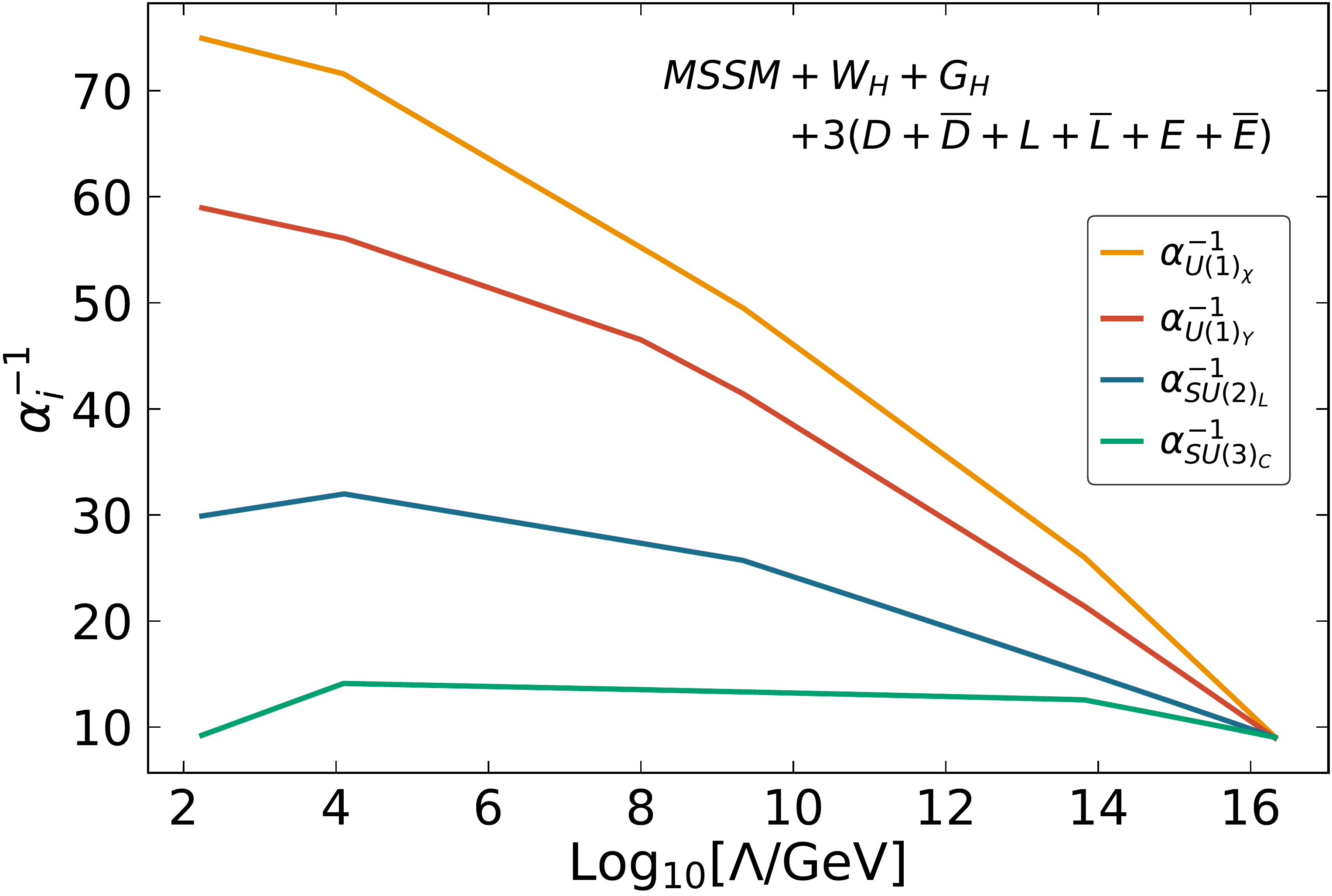}
	\centering \includegraphics[width=8.00cm]{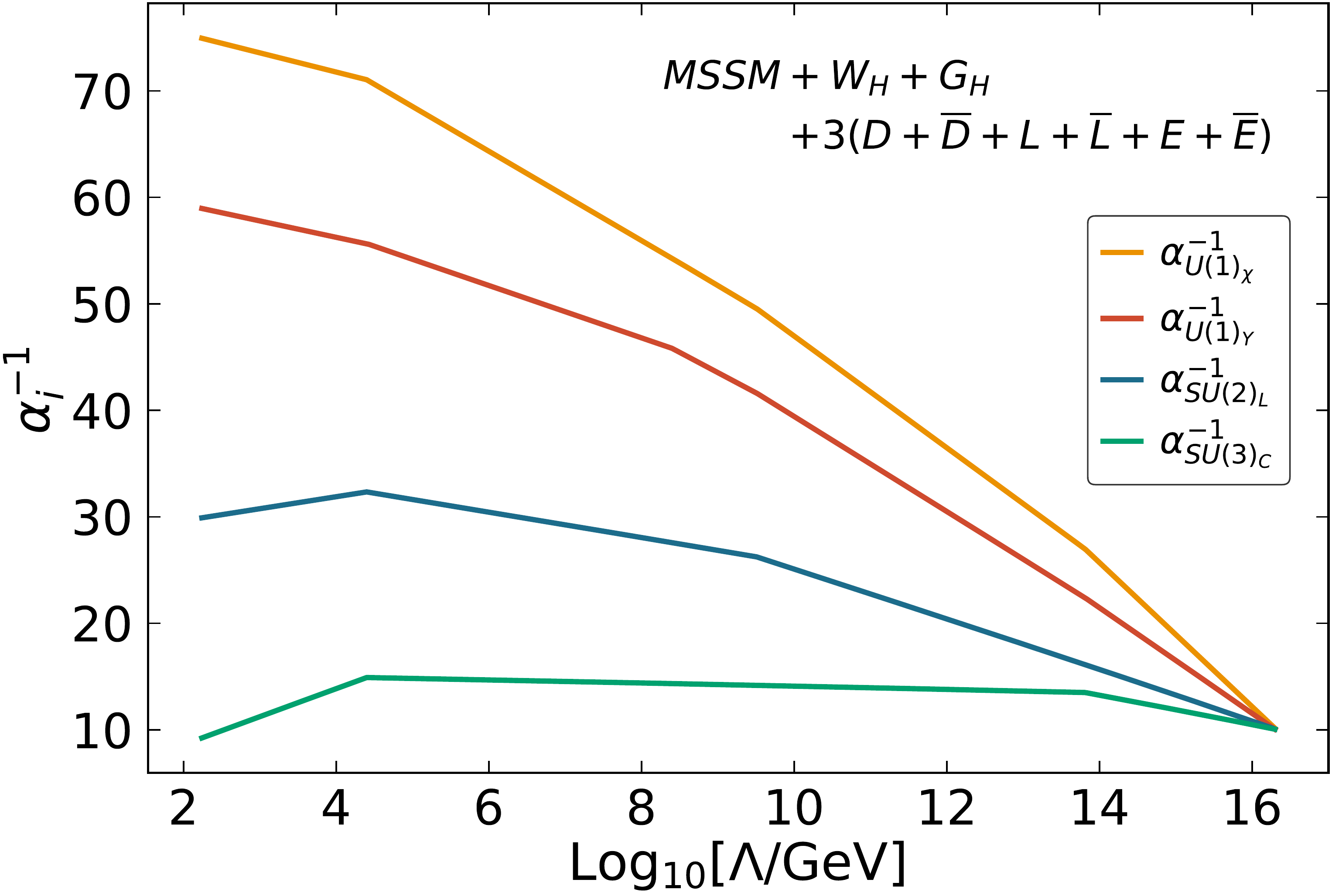}
	\caption{The evolution of the inverse gauge couplings $\alpha_i^{-1}$ with the energy scale $\Lambda$ in $U(1)_R$ symmetric $SU(5) \times U(1)_{\chi}$ model, with SUSY breaking scale $M_{\text{SUSY}} = 12.5$ TeV (left) and $M_{\text{SUSY}} = 25$ TeV (right). Unification is achieved with three generations of vectorlike fermions and GUT scale at $M_{\text{GUT}} \sim 2 \times 10^{16}$ GeV in both cases.}
	\label{fig:unification}
\end{figure}  
The masses for $E + \overline{E}$ and $L + \overline{L}$ can be made light with fine tuning on the parameters such that 
\begin{equation}
	40 \mathcal{A}^{(E,\overline{E})} + 12 \mathcal{B}^{(E,\overline{E})} \sim 0, \qquad
	20 \mathcal{A}^{(5,\overline{5})} + 6 \mathcal{B}^{(5,\overline{5})} \sim 0.
\end{equation} 
The mass of $D + \overline{D}$ component is then given as
\begin{equation}
	M_D \sim \frac{21 \mathcal{A}^{(5,\overline{5})}}{81 \alpha^2} \left( \frac{M^{2}}{m_{P}}\right) .
\end{equation}
\begin{table}[t]
	\setlength\extrarowheight{5pt}
	\centering
	\begin{tabular}{ccccc}
		\hline \hline \rowcolor{Gray}
		& \multicolumn{3}{c}{ Vectorlike Fermion masses (GeV)} &    \\ 
		\rowcolor{Gray}	\multirow{-2}{*}{ \begin{tabular}[c]{@{}c@{}}SUSY breaking \\scale $M_{\text{SUSY}}$\end{tabular} }	    &  $M_D$   &  $M_L$   &  $M_E$ &  \\ \cline{1-5}
		12.5 TeV     &  $6.5 \times 10^{13}$ & $2.14 \times 10^{9}$ & $1.0 \times 10^{8}$ &    \\
		25 TeV   &  $6.5 \times 10^{13}$ & $3.16 \times 10^{9}$ & $2.5 \times 10^{8}$ &  \multirow{-4}{*}{\begin{tabular}[c]{@{}c@{}} Unification scale \\ \\ $M_{\text{GUT}} \simeq 2 \times 10^{16}$ GeV \end{tabular}}  \\ \hline
	\end{tabular}
	\caption{The effective SUSY breaking scales, $M_{\text{SUSY}}$ and corresponding mass splitting patterns of vectorlike families. Unification occurs at the same scale for both cases.}
	\label{tab:vectorlike_mass_splitting}
\end{table}

Fig. \ref{fig:unification} shows successful gauge-coupling unification with three generations of the vectorlike families and different mass-splittings, as listed in Table \ref{tab:vectorlike_mass_splitting}, for two SUSY-breaking scales, $M_{\text{SUSY}} = (12.5,\,25)$ TeV. Here, we assume the masses of the octet and the triplet to be near the SUSY-breaking scale, $M_{\text{SUSY}} \simeq M_{G_H} \simeq M_{W_H}$. In both cases, the unification is achieved at $M_{\text{GUT}} = (5\sqrt{2}/9 \alpha)g_5 M \sim 2 \times 10^{16}$ GeV, where $g_5$, the gauge coupling of $SU(5)$ gauge group, is unified with $g_{\chi}$, the gauge coupling of $U(1)_{\chi}$ group.

\subsection{\large{\bf Dimension-5 Proton Decay}}\label{sec4}

In this section, the implementation of the douplet-triplet solution to the well known issue of the color triplets $D_h, \bar{D}_{\bar{h}}$ embedded in the same representations $\mathbf{5}$ and $\bar{\mathbf{5}}$ with the MSSM Higgs fields is briefly discussed. 
The relevant superpotential terms are 
\be
W \supset \gamma \bar{h} \Phi h + \delta \bar{h} h~.
\ee
After the $SU(5)$ symmetry breaking, these can be written in terms of the MSSM fields as follows
\be
W \supset \left( \delta - \frac{3 \gamma \phi_{24}^0}{2\sqrt{15}} \right) h_u h_d + \left(\delta + \frac{\gamma \phi_{24}^0}{\sqrt{15}} \right) \bar{D}_{\bar{h}} D_h \supset \mu_H h_u h_d + M_{D_h} \bar{D}_{\bar{h}} D_h~.
\ee
We observe that the doublet-triplet splitting problem is resolved by requiring fine-tuning of the involved parameters, 
such that
$$\delta \sim \frac{3 \gamma \phi_{24}^0}{2\sqrt{15}} ~.$$
Here $\mu_H$ is identified with the MSSM $\mu$ parameter taken to be of the order of TeV scale while, $M_{D_h}$ is the color triplet Higgs mass parameter given by,
\begin{equation}
	M_{D_h} \sim \frac{5 \, \gamma \phi_{24}^0}{2\sqrt{15}} = \frac{5 \gamma M_{\alpha} y_2}{\sqrt{30 \left( \frac{4}{27 \alpha^2} -1 \right)}}.
\end{equation}  
The dominant contribution to dimension-5 proton decay amplitude comes from color-triplet Higgsinos and typically dominates the decay rate from gauge boson mediated dimension-6 operators. The proton lifetime for the decay $p \rightarrow K^+ \bar{\nu}$ mediated by color-triplet Higgsinos is approximated by \cite{Nagata:2013ive}:
\begin{equation}
	\tau_p \simeq 4 \times 10^{35} \times \sin^4 2\beta \left( \frac{M_{\text{SUSY}}}{10^2 ~ \text{TeV}} \right)^2 \left( \frac{M_{D_h}}{10^{16} ~ \text{GeV}} \right)^2 \text{years}, \label{proton_lifetime}
\end{equation}
which depends on Higgino mass as well as the SUSY breaking scale $M_{\text{SUSY}}$. The Super-Kamiokande experiment places a lower bound on proton lifetime of $\tau_p = 5.9 \times 10^{33}$ years at $90\%$ confidence level for the channel $p \rightarrow K^+ \bar{\nu}$. With $M_{\alpha} \simeq 2 \times 10^{16}$ GeV, this translates into a lower bound on $M_{\text{SUSY}}$,
\begin{equation}
	M_{\text{SUSY}} \gtrsim 12.5 ~ \text{TeV} .
\end{equation}
\begin{figure}[t]
	\centering
	\includegraphics[width=0.6\textwidth]{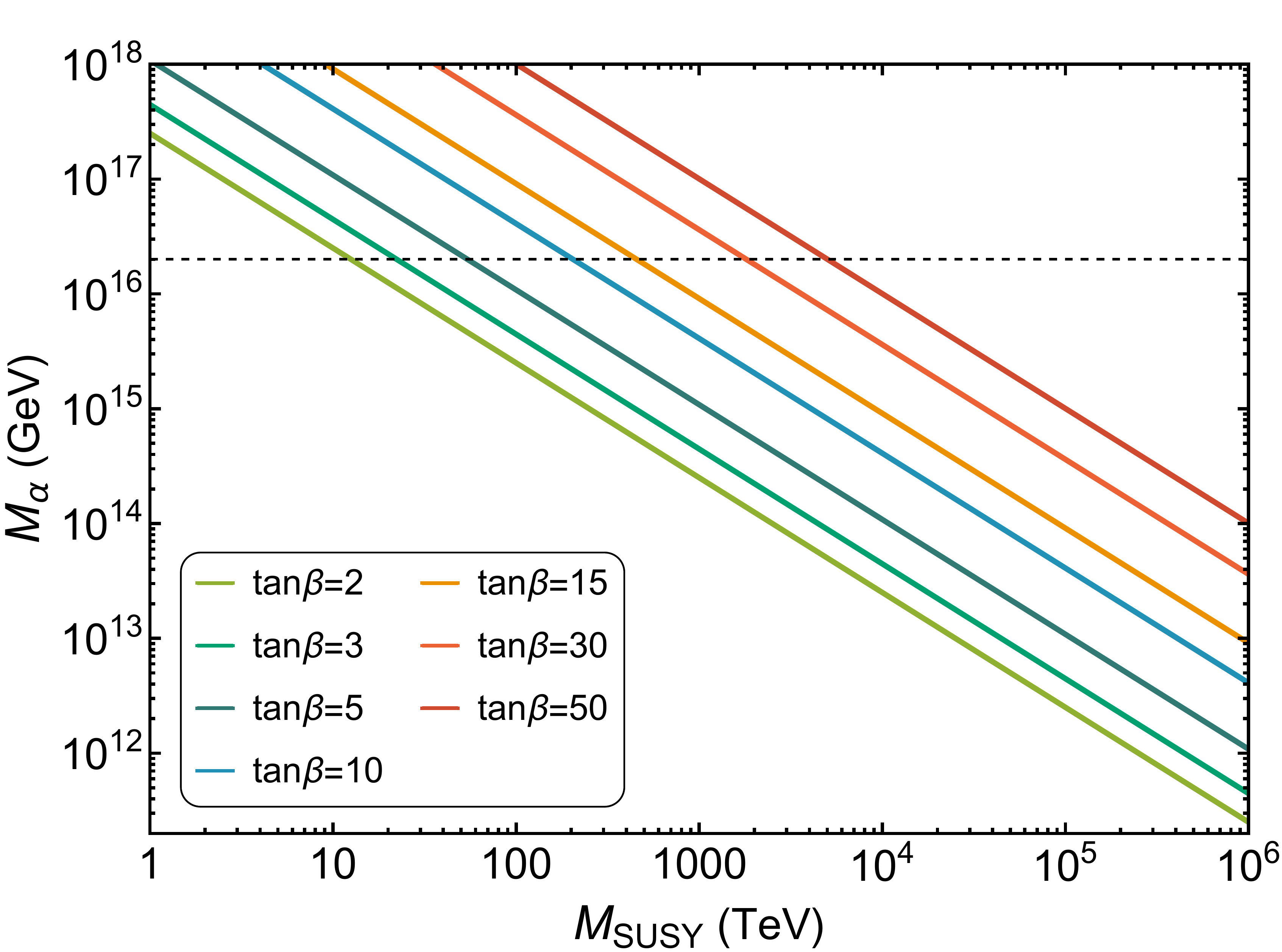}
	\label{fig:superk_bound}
	\caption{$SU(5)$ gauge symmerty breaking scale $M_{\alpha}$ as a function of SUSY breaking scale $M_{\text{SUSY}}$ for different values of $\tan \beta$. The curves are drawn for proton lifetime fixed at Super-Kamiokande bounds ($\tau_p = 5.9 \times 10^{33}$ years).}
	\label{fig:proton_decay}
\end{figure}
This can also be seen in Fig. \ref{fig:proton_decay} where $SU(5)$ gauge symmerty breaking scale $M_{\alpha}$ is plotted against the SUSY breaking scale $M_{\text{SUSY}}$. The curves represent different values of $\tan \beta$ and drawn for proton lifetime fixed at Super-Kamiokande bounds \cite{Super-Kamiokande:2016exg}.

\section{\large{\bf Minimal K\"ahler Potential}}\label{sec5}

The minimal canonical K\"ahler potential is given as,
\be \label{Ktree}
K = \vert S \vert^2 + Tr \vert \Phi \vert^2
+ \vert h \vert^2 + \vert \bar{h}\vert^2 + \vert \chi \vert^2 + \vert \bar{\chi} \vert^2+ \vert \nu_{i}^{c} \vert^2 .
\ee
The F-term SUGRA scalar potential is given by 
\begin{equation}
	V_{\text{SUGRA}}=e^{K/m_P^{2}}\left(
	K_{i\bar{j}}^{-1}D_{z_{i}}WD_{z^{*}_j}W^{*}-3 m_P^{-2}\left| W\right| ^{2}\right),
	\label{VF}
\end{equation}
with $z_{i}$ being the bosonic components of the superfields $z%
_{i}\in \{S,\Phi,h,\bar{h}, \chi, \bar{\chi} ,\cdots\}$, and we have defined
\be
D_{z_{i}}W \equiv \frac{\partial W}{\partial z_{i}}+m_P^{-2}\frac{%
	\partial K}{\partial z_{i}}W , \,\,\,
K_{i\bar{j}} \equiv \frac{\partial ^{2}K}{\partial z_{i}\partial z_{j}^{*}},
\ee
and $D_{z_{i}^{*}}W^{*}=\left( D_{z_{i}}W\right)^{*}.$
The SUGRA scalar potential during inflation becomes
\begin{eqnarray}
\begin{split}
	V_{\text{SUGRA}}&=\kappa ^2 M_{\alpha}^4 \left[1+ \left(\frac{4}{9\left(4/27 - \alpha^2\right)}\right)\left(\frac{M_{\alpha }}{m_{p}}\right)^2 +\right.\\
&\left. \left(\frac{4 \left(2 + 9 x^2\left(4/27 - \alpha^2 \right) \right)}{81\left(4/27 - \alpha^2\right)^2} + \frac{1}{2} x^4\right)\left(\frac{M_{\alpha }}{m_{p}}\right)^4+.....\right]\, .
\end{split}
\end{eqnarray}
Putting all these corrections together, we obtain the following form of inflationary potential,
\begin{eqnarray}\nonumber
	V &\simeq& V_{\text{SUGRA}} + V_{\text{1-loop}} + V_{\text{Soft}} \\ \nonumber
	&\simeq& \kappa ^2 M_{\alpha}^4 \Bigg[1+ \left(\frac{4}{9\left(4/27 - \alpha^2\right)}\right)\left(\frac{M_{\alpha }}{m_{p}}\right)^2 \\ \nonumber
	&+&  \left(\frac{4 \left(2 + 9 x^2\left(4/27 - \alpha^2 \right) \right)}{81\left(4/27 - \alpha^2\right)^2} + \frac{1}{2} x^4\right)\left(\frac{M_{\alpha }}{m_{p}}\right)^4 \\ \nonumber
	&+& \frac{\kappa^2}{16 \pi^2} \left[ F(M_{\alpha}^2,x^2) 	+ 11\times 25\,F(5 M_{\alpha}^2,5\,x^2) \right] + \frac{\sigma_{\chi}^2}{8 \pi^2} F(M_{\alpha}^2,y^2) \\ 
	&+&  \frac{a m_{3/2} x}{\kappa M_{\alpha}}  + \frac{M_S^2\, x^2}{\kappa^2 M_{\alpha}^2}     + \frac{8 M_{\phi}^2}{9 \kappa^2 M_{\alpha}^2 \left( 4/27 - \alpha^2 \right)} \Bigg] .
\end{eqnarray}
The inflationary slow roll parameters are given by,
\bea
\epsilon = \frac{1}{4}\left( \frac{m_P}{M_{\alpha}}\right)^2
\left( \frac{V'}{V}\right)^2, \,\,\,
\eta = \frac{1}{2}\left( \frac{m_P}{M_{\alpha}}\right)^2
\left( \frac{V''}{V} \right), \,\,\,
\alpha^2 = \frac{1}{4}\left( \frac{m_P}{M_{\alpha}}\right)^4
\left( \frac{V' V'''}{V^2}\right).
\eea
Here, the derivatives are with respect to $x=|S|/M_{\alpha}$, whereas the canonically normalized field $\sigma \equiv \sqrt{2}|S|$. In the slow-roll (leading order) approximation, the tensor-to-scalar ratio $r$, the scalar spectral index $n_s$, and the running of the scalar spectral index $dn_s / d \ln k$ are given by
\bea
r &\simeq& 16\,\epsilon,  \\
n_s &\simeq& 1+2\,\eta-6\,\epsilon,  \\
\frac{d n_s}{d\ln k} &\simeq& 16\,\epsilon\,\eta
-24\,\epsilon^2 - 2\,\xi^2 .
\eea
\begin{figure}[t]
	\centering \includegraphics[width=8.035cm]{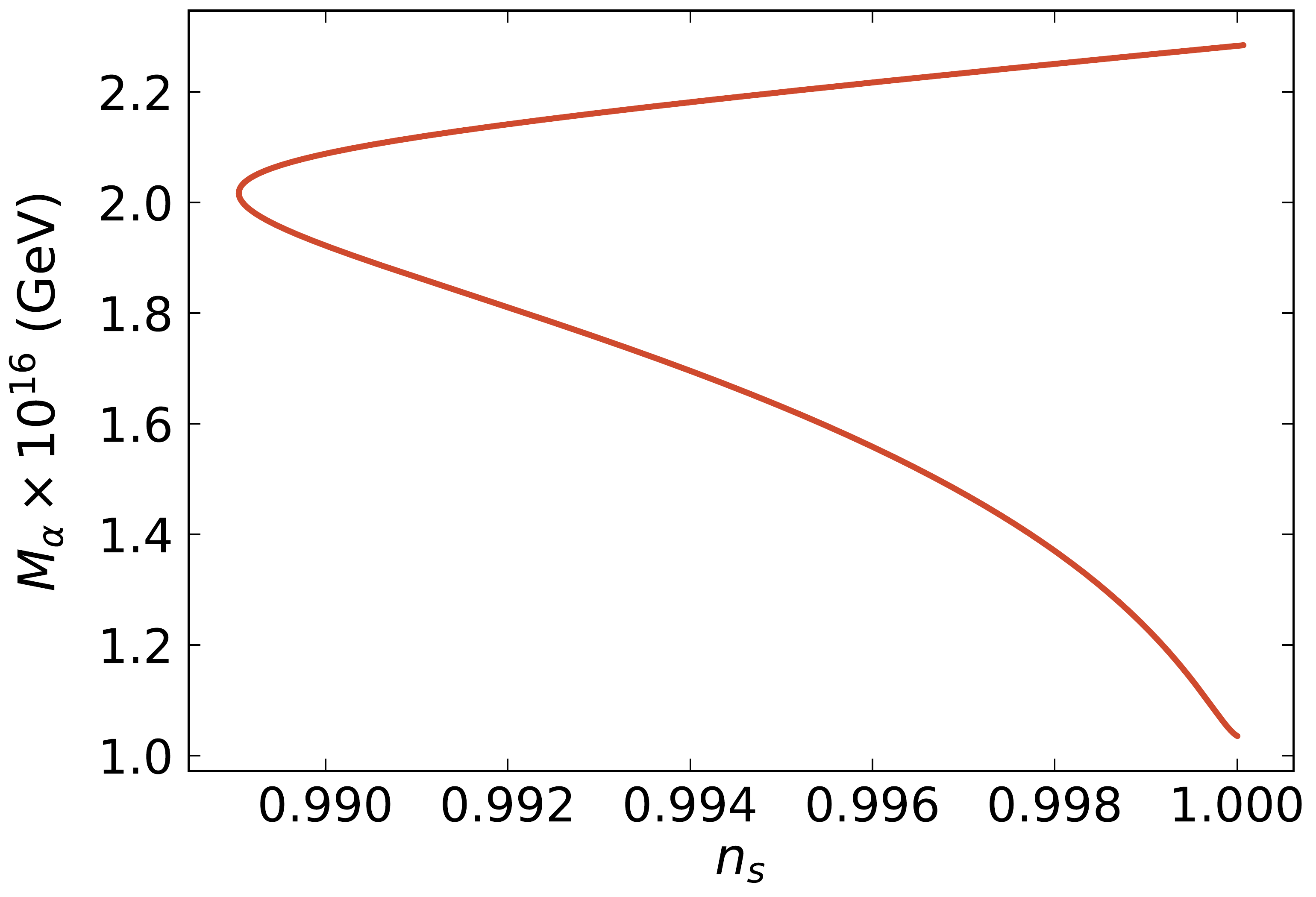}
	\centering \includegraphics[width=8.035cm]{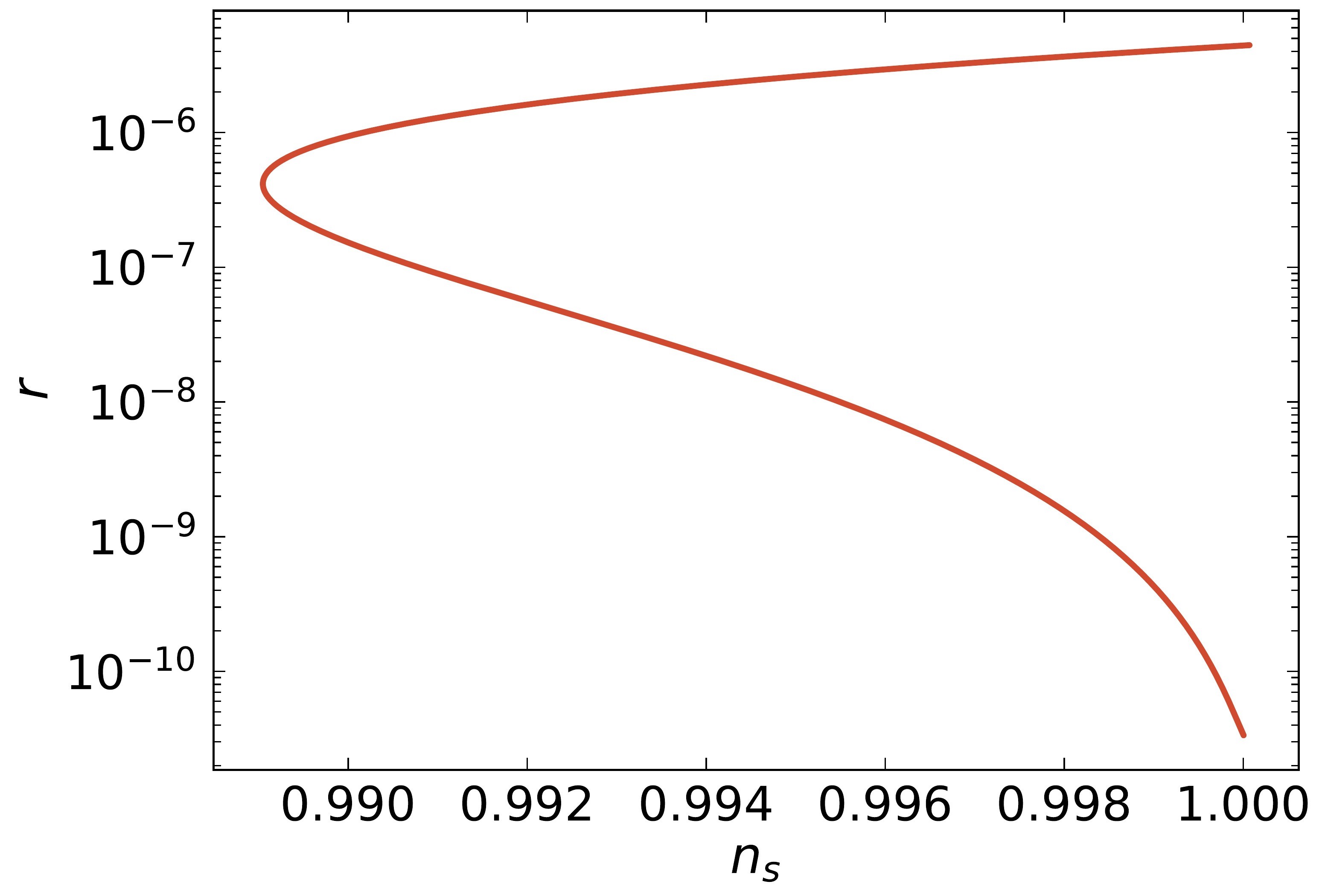}
	\centering \includegraphics[width=8.035cm]{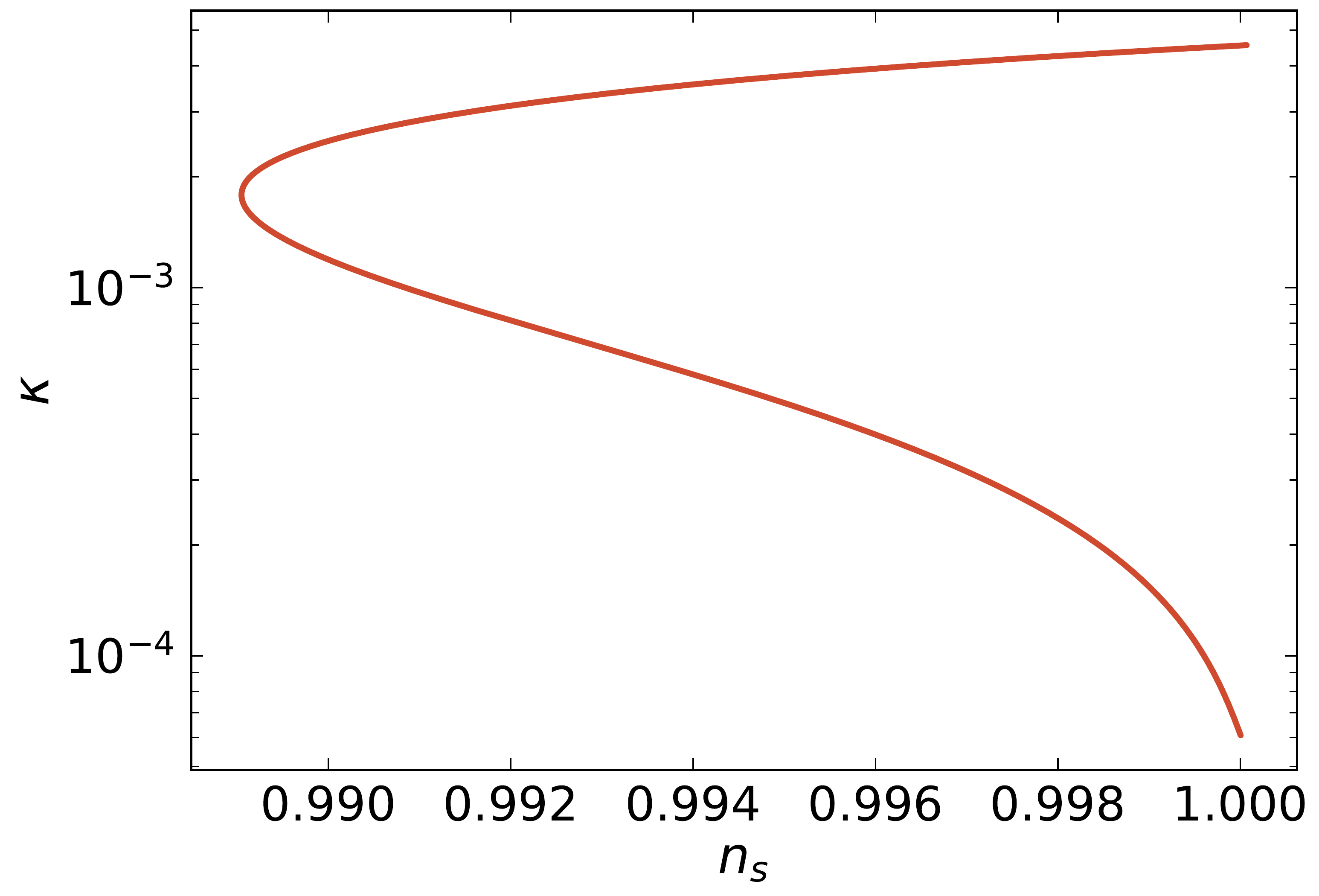}
	\centering \includegraphics[width=8.035cm]{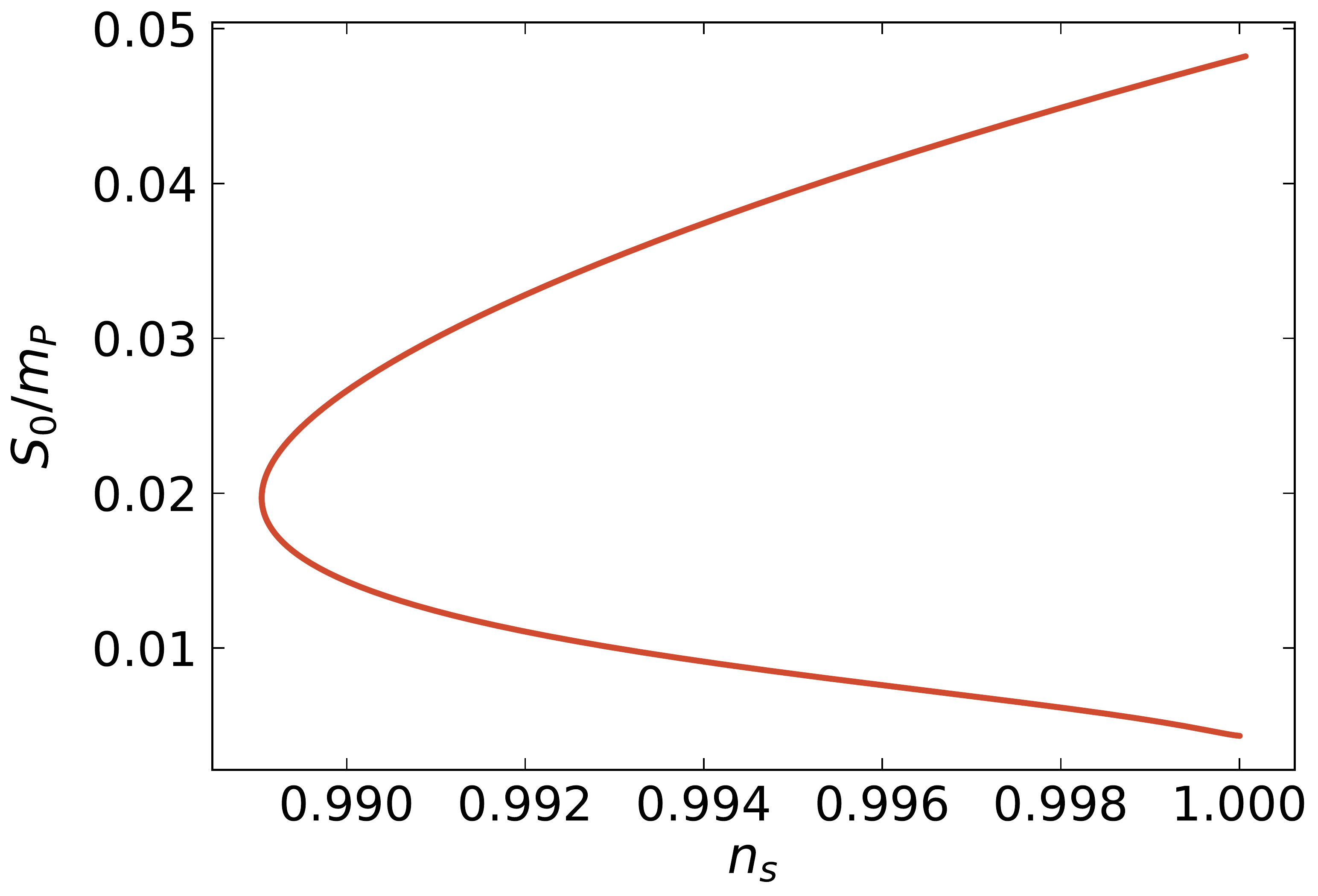}
	\caption{The scalar spectral index $n_s$ vs the $SU(5)$ symmetry breaking scale $M_{\alpha}$, the tensor-to-scalar ratio $r$, $\kappa$ and $S_0/m_P$ for minimal K\"ahler potential without the soft mass terms.}
	\label{fig:minimal_nosoft}
\end{figure}
\begin{figure}[t]
	\centering \includegraphics[width=8.035cm]{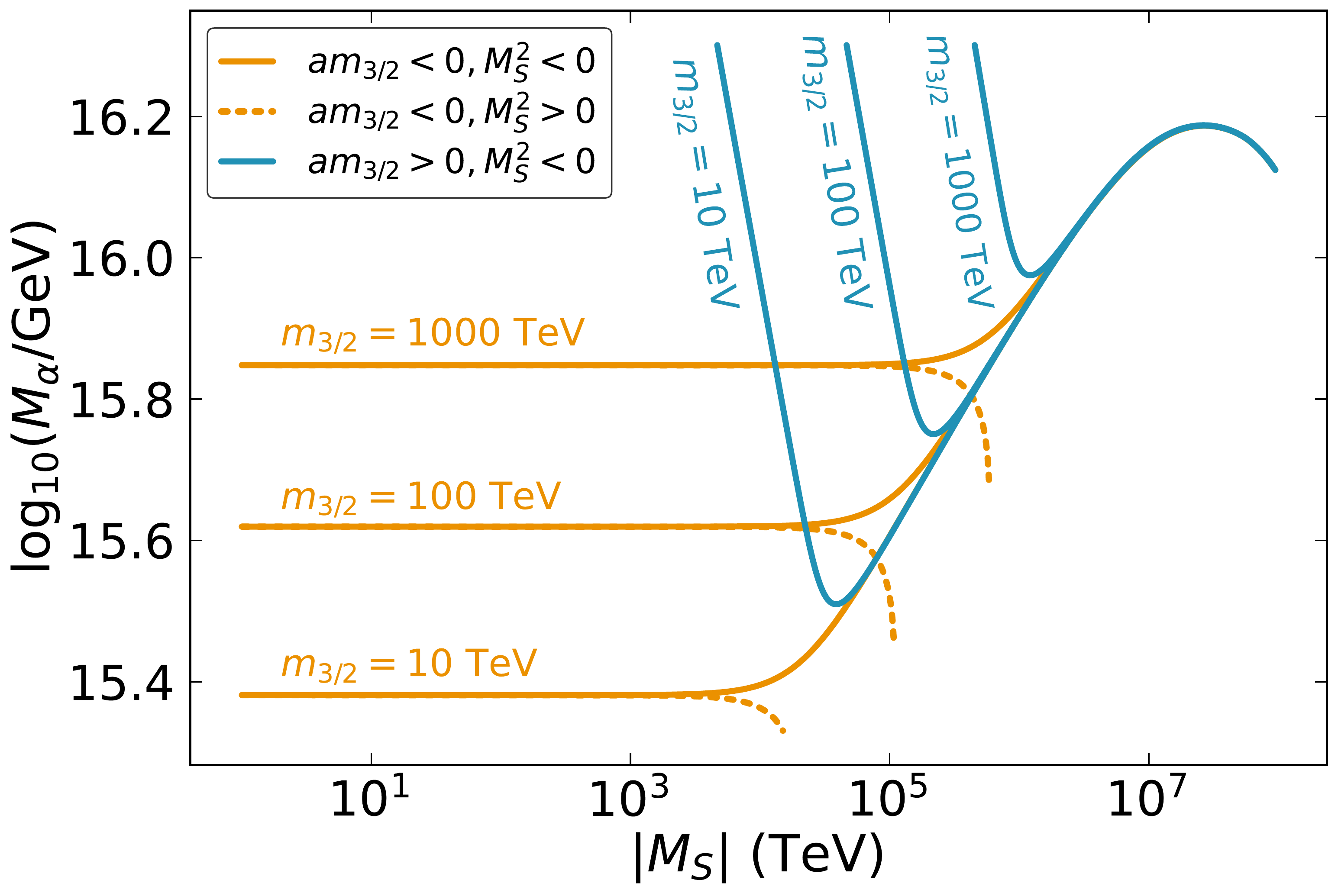}
	\centering \includegraphics[width=8.035cm]{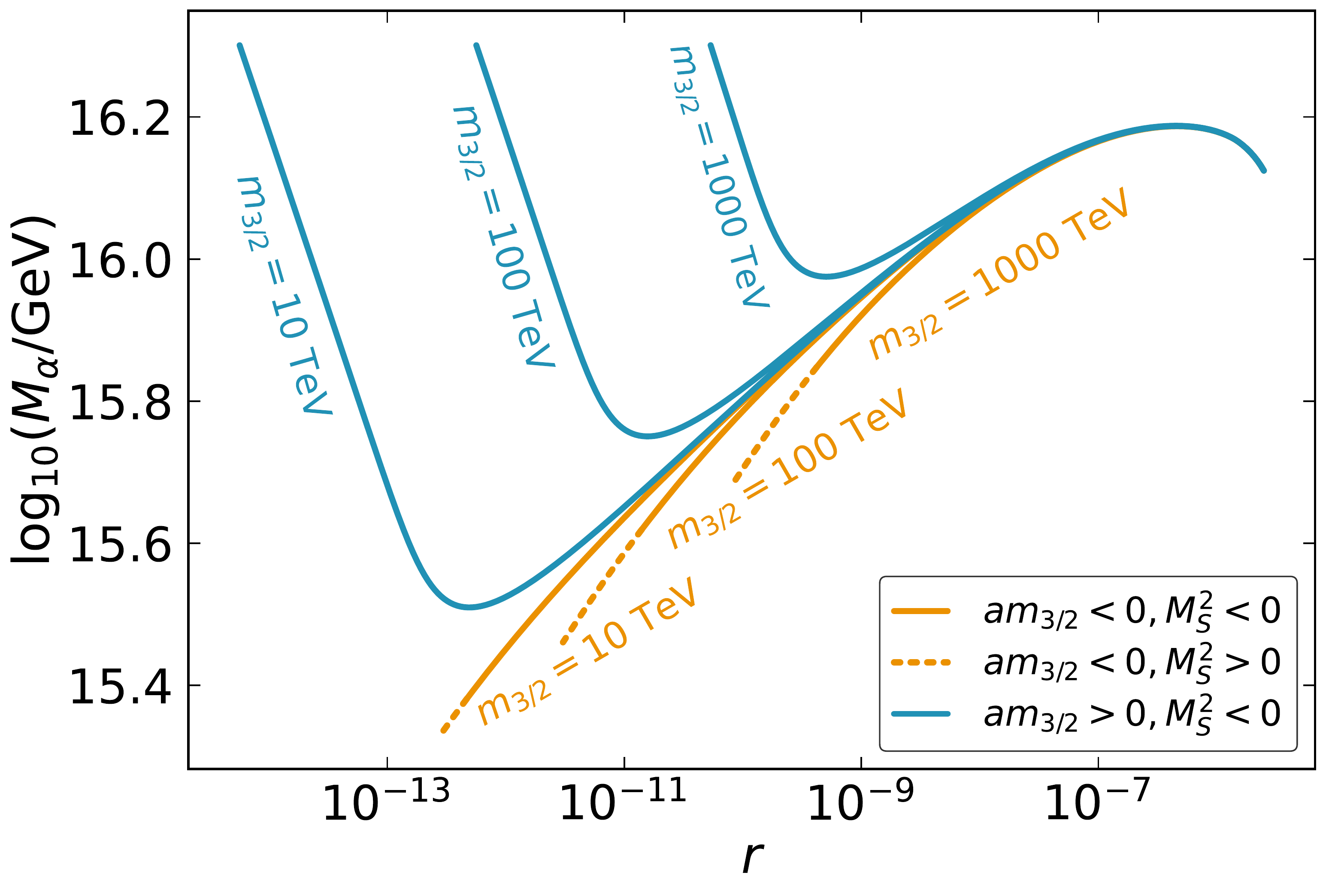}
	\centering \includegraphics[width=8.035cm]{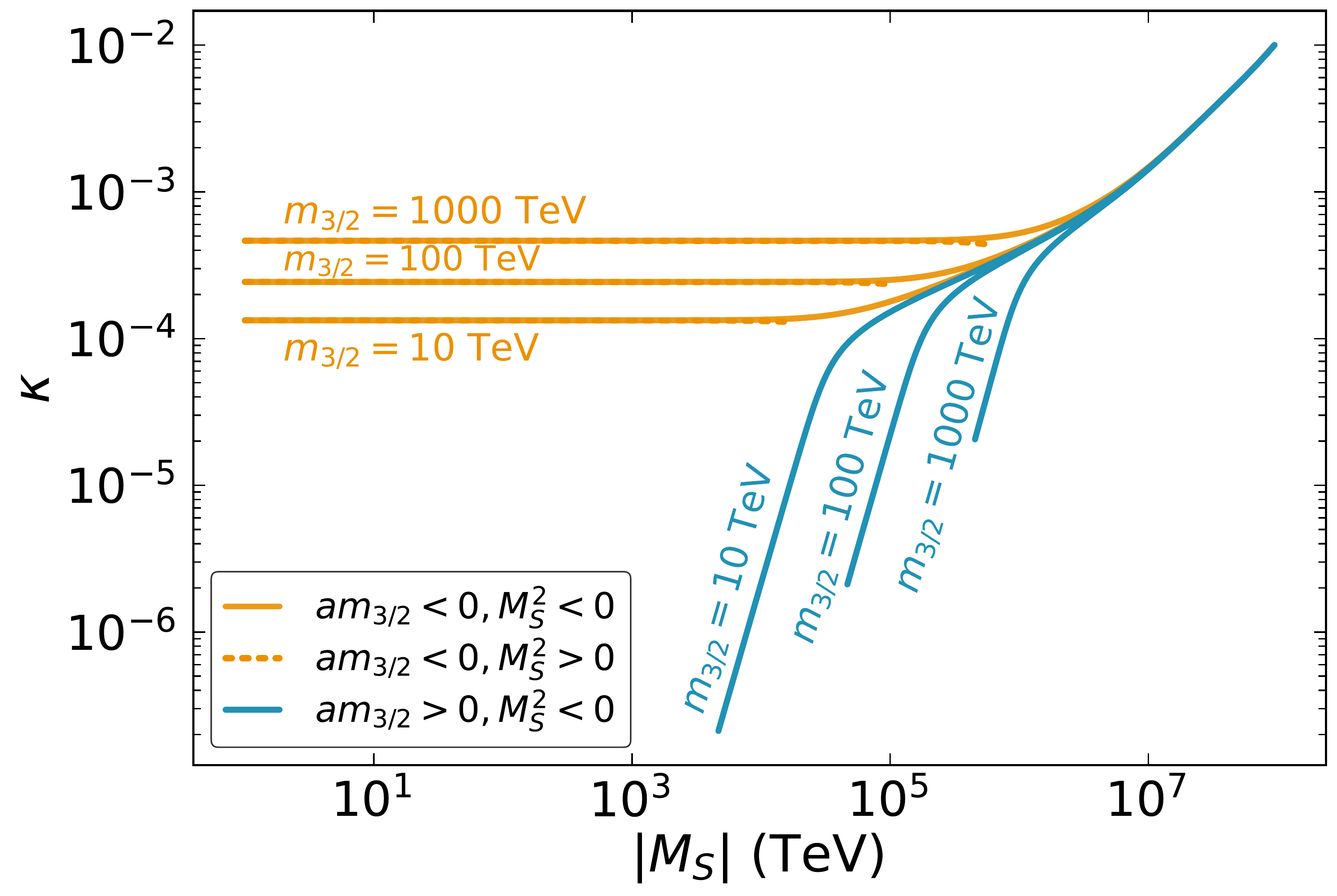}
	\centering \includegraphics[width=8.035cm]{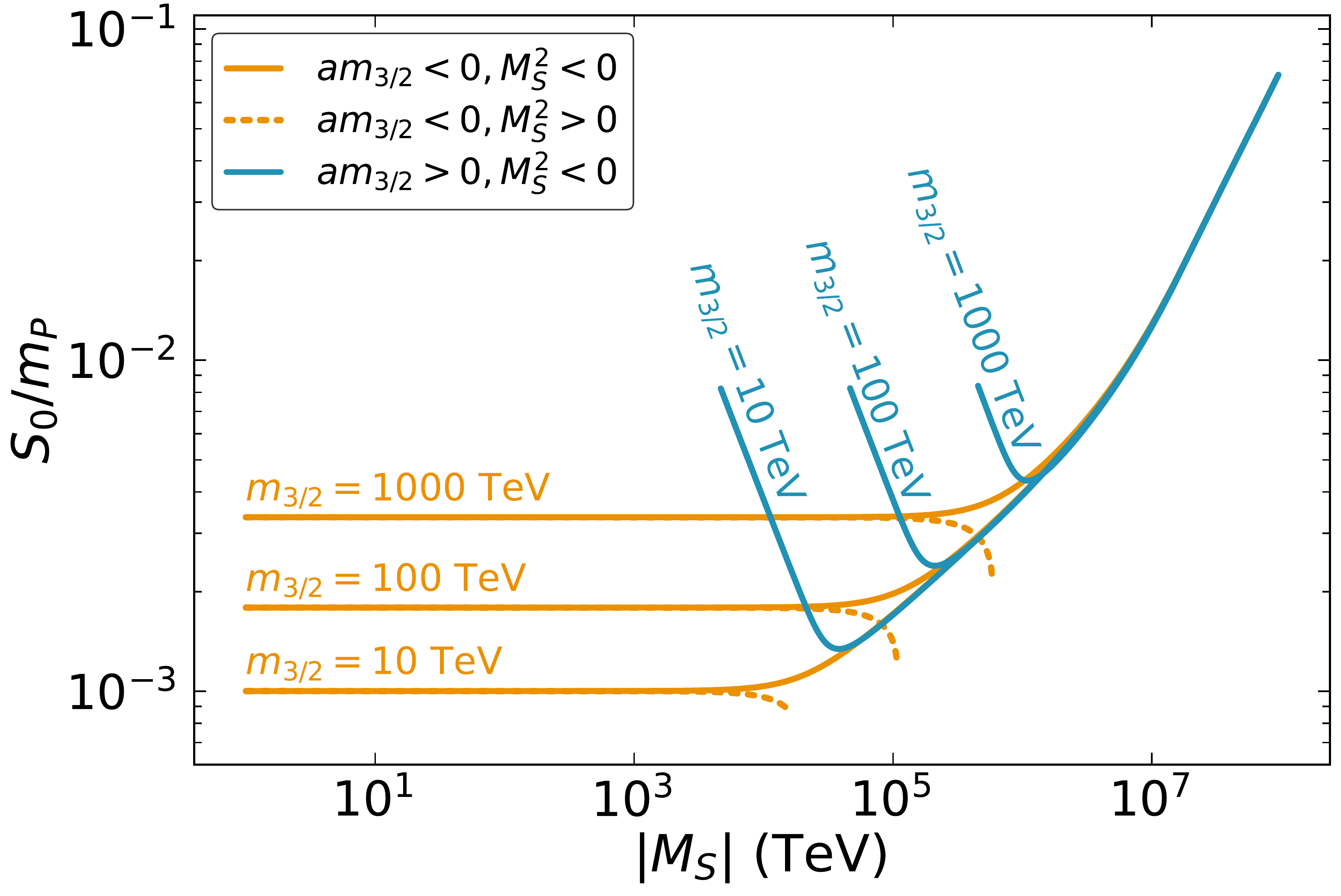}
	\caption{The scalar spectral index $n_s$ vs the $SU(5)$ symmetry breaking scale $M_{\alpha}$, the tensor-to-scalar ratio $r$, $\kappa$ and $S_0/m_P$ for minimal K\"ahler potential without the soft mass terms.}
	\label{fig:minimal_soft}
\end{figure}
The last $N_0$ number of e-folds before the end of inflation is,
\bea
N_0 = 2\left( \frac{M_{\alpha}}{m_P}\right) ^{2}\int_{x_e}^{x_{0}}\left( \frac{V}{%
	V'}\right) dx,
\label{efolds}
\eea
where $x_0$ is the field value at the pivot scale $k_0$, and
$x_e$ is the field value at the end of inflation, defined by $|\eta(x_e)| = 1$.  Assuming a standard thermal history, $N_0$ is related to $T_r$ as \cite{Garcia-Bellido:1996egv}
\begin{equation}\label{n0}
	N_0=54+\frac{1}{3}\ln\Big(\frac{T_r}{10^9\text{ GeV}}\Big)+\frac{2}{3}\ln\Big(\frac{V(x)^{1/4} }{10^{15}\text{ GeV}}\Big),
\end{equation}
where $T_r$ is the reheat temperature and in numerical calculation we set $T_r=10^9$ GeV. This could easily be reduced to lower values if the gravitino problem is regarded to be an issue.\footnote{For a recent discussion on the gravitino overproduction problem in hybrid inflation see Ref.\cite{Nakayama:2010xf}} The amplitude of the curvature perturbation is given by \cite{Liddle:1993fq}
\bea
A_{s}(k_0) = \frac{1}{24\,\pi^2}
\left. \left( \frac{V/m_P^4}{\epsilon}\right)\right|_{x = x_0},
\label{perturb}
\eea
where $A_{s}= 2.137 \times 10^{-9}$ is the Planck normalization at $k_0 = 0.05\, \rm{Mpc}^{-1}$. In our numerical calculations, we have taken $M_S = M_{\phi}$ and have set the dimensionless couplings equal, $\kappa = \sigma_{\chi}$, such that $\zeta = 1$. Fig. \ref{fig:minimal_nosoft}, shows our results without soft SUSY mass terms where various parameters are plotted against the scalar spectral index $n_s$. It can be seen that without the soft mass terms, the scalar spectral index $n_s$ cannot be achieved within Plank 2-$\sigma$ bounds. With the inclusion of soft mass terms, the scalar spectral index $n_s$ can easily be obtained within Planck's 2-$\sigma$ bounds. 

The soft mass terms, seem to play an important role in inflationary predictions. Fig. \ref{fig:minimal_soft} shows our numerical results with soft mass terms where the behavior of $SU(5)$ guage symmetry breaking scale $M_{\alpha}$ (upper left panel), $\kappa$ (lower left panel) and $S_0/m_P$ (lower right panel) is depicted as a function of soft mass parameter $\vert M_S \vert $ for different values of the gravitino mass $m_{3/2}$. The behavior of $SU(5)$ guage symmetry breaking scale $M_{\alpha}$ with respect to the tensor to scalar ratio $r$ is shown in the upper right panel. In obtaining these results, we have fixed the scalar spectral index $n_s$ at the central value (0.9655) of Planck's latest bounds. The soft mass squared parameter $M_S^2$ and the combination $a m_{3/2}$ can be either positive or negative. We consider the following possible cases in our numerical calculations,
\begin{gather}\nonumber
	a m_{3/2} > 0 \quad M_S^2 > 0 ,\\ \nonumber
	a m_{3/2} < 0 \quad M_S^2 < 0 ,\\ \nonumber
	a m_{3/2} < 0 \quad M_S^2 > 0 ,\\ 
	a m_{3/2} > 0 \quad M_S^2 < 0 .
\end{gather}
The first case with $M_S^2 > 0$ and $a m_{3/2} > 0$ ($a = +1$) is inconsistent with Planck's results. A red tilted scalar spectral index ($n_s < 1$) compatible with Planck's latest bounds is obtained for the rest of the cases. The yellow curves are drawn for $a m_{3/2} < 0$ ($a = -1$) where the solid lines correspond to the case when $M_S^2 < 0$ and dashed lines correspond to $M_S^2 > 0$. For lower values of $\vert M_S \vert \simeq (1 - 10^4)$ TeV, the radiative corrections provide dominant contribution, whereas both SUGRA corrections and soft mass squared terms are suppressed.  The suppression of supergravity corrections in this region is supported by small values of $S_0/m_P \simeq 10^{-3}$, as shown in lower right panel of Fig. \ref{fig:minimal_soft}. For $\vert M_S \vert  \gtrsim 10^4$ TeV, the soft mass squared term begins to take over, which drives the curve upward for $M_S^2 < 0$, and downward for $M_S^2 > 0$. For $M_S^2 < 0$ and $\vert M_S \vert  \gtrsim 10^4$ TeV, $\kappa$ takes on large values, $M_{\alpha}$ approaches $\sim 1.5 \times 10^{16}$ GeV and supergravity corrections become important. 

It is useful to analytically examine some approximate equations to understand the behavior depicted in Fig. \ref{fig:minimal_soft}.  In the slow-roll approximation, the amplitude of the power spectrum of scalar curvature perturbation $A_{s}$ and the scalar spectral index $n_s$ is given by,
\begin{eqnarray}\label{cur_analytic}
	A_{s}(k_{0}) &\simeq& \frac{\kappa^2}{6\,\pi^2}\left(\frac{M_{\alpha}}{m_{p}}\right)^{6} \left( 2 x_0^3\left(\frac{M_{\alpha}}{m_P}\right)^4 - \frac{2 M_S^2 x_0}{\kappa^2 M_{\alpha}^2} + \frac{a m_{3/2}}{\kappa M_{\alpha}} + \frac{278 \kappa^2}{16\,\pi^2} F^{'}(5 x_0) \right)^{-2}, \\ \label{ns_analytic}
	n_s &\simeq& 1 +  \left(\frac{m_{p}}{M_{\alpha}}\right)^{2} \left( 6 x_0^2\left(\frac{M_{\alpha}}{m_P}\right)^4 - \frac{2 M_S^2}{\kappa^2 M_{\alpha}^2} + \frac{278 \kappa^2}{16\,\pi^2}  F^{''}(5 x_0) \right) .
\end{eqnarray}
Taking the contributions of the soft linear mass term comparable to mass squared term, we obtain the following analytical expressions for $M_{\alpha}$, $\kappa$ and $\vert M_S \vert$;

\begin{gather}
	\kappa \simeq \left( \left( \frac{8 \pi^3}{139} \right)^3 \frac{\left(1 - n_s\right)}{\vert F^{'} (5 x_0) \vert^2  \vert F^{''} (5 x_0) \vert}  \right)^{1/8} \left( \frac{m_{3/2}}{m_P} \right)^{1/4}, \\
	M_{\alpha} \simeq \left(  \frac{139 \vert F^{''} (5 x_0) \vert^3}{8 \pi^2 \left( 1 - n_s \right)^3 \vert F^{'} (5 x_0) \vert^2} \right)^{1/8} \left( m_{3/2} m_P^3 \right)^{1/4} , \\
	\vert M_S \vert \simeq \kappa^2 M_{\alpha} \sqrt{\frac{139}{32 \pi^2} F^{'} (5 x_0)} \,.
\end{gather}
It can be checked that, for $m_{3/2} \simeq 10$ TeV, $n_s = 0.9655$, $x_0 \sim 1$, we obtain $\kappa \simeq 1.3 \times 10^{-4}$, $M_{\alpha} \simeq 2 \times 10^{15}$ GeV and $\vert M_S \vert \simeq 2 \times 10^4$ TeV. Also, for $m_{3/2} \simeq 1000$ TeV, $n_s = 0.9655$, $x_0 \sim 1$, we obtain $\kappa \simeq 4.5 \times 10^{-4}$, $M_{\alpha} \simeq 5.5 \times 10^{15}$ GeV and $\vert M_S \vert \simeq 7 \times 10^5$ TeV. These estimates are in excellent agreement with the numerical results shown in Figs. \ref{fig:minimal_soft}. Therefore, for $m_{3/2} \lesssim 1000$ TeV and $M_S \lesssim 10^4$ TeV, only radiative corrections and linear soft mass term dominate, whereas the SUGRA corrections and soft mass-squared term are suppressed. It should be noted that larger values of $m_{3/2}$ shift the contribution of the soft mass-squared term towards larger values of $M_S$. For example, with $m_{3/2} \simeq 10$ TeV, the soft mass-squared term begins to take over for $M_S \gtrsim 10^4$ TeV, whereas with $m_{3/2} \simeq 1000$ TeV, the soft mass-squared term becomes important for $M_S \gtrsim 5 \times 10^5$ TeV. Furthermore, for larger values of $m_{3/2}$ ($\gtrsim 1000$ TeV), the SUGRA corrections become important and large values of $M_{\alpha}$ can be obtained, independent of $M_S$. The curves exhibit similar behavior for $\vert M_S \vert \gtrsim 10^6$ TeV for all three cases but decouple for value of $\vert M_S \vert$ below $10^6$ TeV.

The fourth case with $a m_{3/2} > 0$ ($a = +1$) and $M_S^2 <0$, generates large $M_{\alpha}$ that easily approaches $M_{\text{GUT}}$. For $\vert M_S \vert \lesssim 10^6$ TeV, $M_{\alpha}$ takes on large values, whereas the radiative corrections become suppressed owing to small values of $\kappa$. This is in contrast to the other two cases where the contribution of soft mass-squared term becomes negligible below $\vert M_S \vert \lesssim 10^6$ TeV. With radiative corrections suppressed, Eqs. \eqref{cur_analytic} and \eqref{ns_analytic} simplify to the following form,
\begin{eqnarray}\label{cur_analytic_2}
	A_{s}(k_{0}) &\simeq& \frac{\kappa^2}{6\,\pi^2}\left(\frac{M_{\alpha}}{m_{p}}\right)^{6} \left( 2 x_0^3\left(\frac{M_{\alpha}}{m_P}\right)^4 - \frac{2 M_S^2 x_0}{\kappa^2 M_{\alpha}^2} + \frac{a m_{3/2}}{\kappa M_{\alpha}} \right)^{-2}, \\ \label{ns_analytic_2}
	n_s &\simeq& 1 +  \left(\frac{m_{p}}{M_{\alpha}}\right)^{2} \left( 6 x_0^2\left(\frac{M_{\alpha}}{m_P}\right)^4 - \frac{2 M_S^2}{\kappa^2 M_{\alpha}^2}  \right) .
\end{eqnarray}
Taking the soft mass-squared term to be comparable to linear soft SUSY-breaking term, we obtain the following analytical expressions for $\kappa$ and $M$ in terms of $m_{3/2}$ and $M_S$;
\begin{eqnarray}
	\kappa \simeq 2 \left(2 \left(1 - n_s \right)\right)^{1/2} \frac{\vert M_S \vert^3}{m_{3/2} m_P} \, ,\\
	M \simeq \left( \frac{1}{2 \left(1 - n_s\right)} \right)^{1/2} \left( \frac{m_{3/2} m_P}{\vert M_S \vert} \right) .
\end{eqnarray}
Using $n_s \simeq 0.9655$, $m_{3/2} \simeq 10$ TeV and $\vert M_S \vert \simeq 4.6 \times 10^3$ TeV, we obtain $\kappa \sim 2 \times 10^{-7}$ and $M_{\alpha} \sim 2 \times 10^{16}$ GeV. Similarly for $n_s \simeq 0.9655$, $m_{3/2} \simeq 1000$ TeV and $\vert M_S \vert \simeq 4.5 \times 10^5$ TeV, we obtain $\kappa \sim 2 \times 10^{-5}$ and $M_{\alpha} \sim 2 \times 10^{16}$ GeV. These estimates are in good agreement with our numerical results displayed in Fig. \ref{fig:minimal_soft}.

The behavior of $M_{\alpha}$ with respect to the tensor to scalar ratio $r$ is shown in the upper right panel of Fig. \ref{fig:minimal_soft} and can be understood from the following approximate relation between $r$, $M_{\alpha}$ and $\kappa$, obtained by using the Planck's normalization constraint on $A_s$,
\begin{equation} \label{eq:rkappaexplicit}
	r \simeq \left( \frac{2 \kappa^2}{3 \pi^2 A_s (k_0)} \right) \left(\frac{M_{\alpha}}{m_P}\right)^4 .
\end{equation}
This shows that $r$ is proportional to both $M_{\alpha}$ and $\kappa$ and large values of $r$ are obtained for large $M_{\alpha}$ and $\kappa$. It can readily be checked that for $M_{\alpha} \simeq 2.4 \times 10^{15}$ GeV and $\kappa \simeq 1.3 \times 10^{-4}$, the above equation gives $r \simeq 4.7 \times 10^{-7}$. Similarly, for $M_{\alpha} \simeq 1.3 \times 10^{16}$ GeV and $\kappa \simeq 0.01$, we obtain $r \simeq 2.5 \times 10^{-6}$. These approximate values are very close to the actual values obtained in our numerical calculations. The above equation therefore gives a valid approximation of our numerical results. The tensor to scalar ratio $r$ varies in the range $4.7 \times 10^{-13} \lesssim r \lesssim 2.5 \times 10^{-6}$ and is beyond the current measuring limits of various upcoming experiments.

\section{\large{\bf Non-Minimal K\"ahler Potential}}\label{sec6}

\begin{figure}[!htb]
	\centering \includegraphics[width=8.035cm]{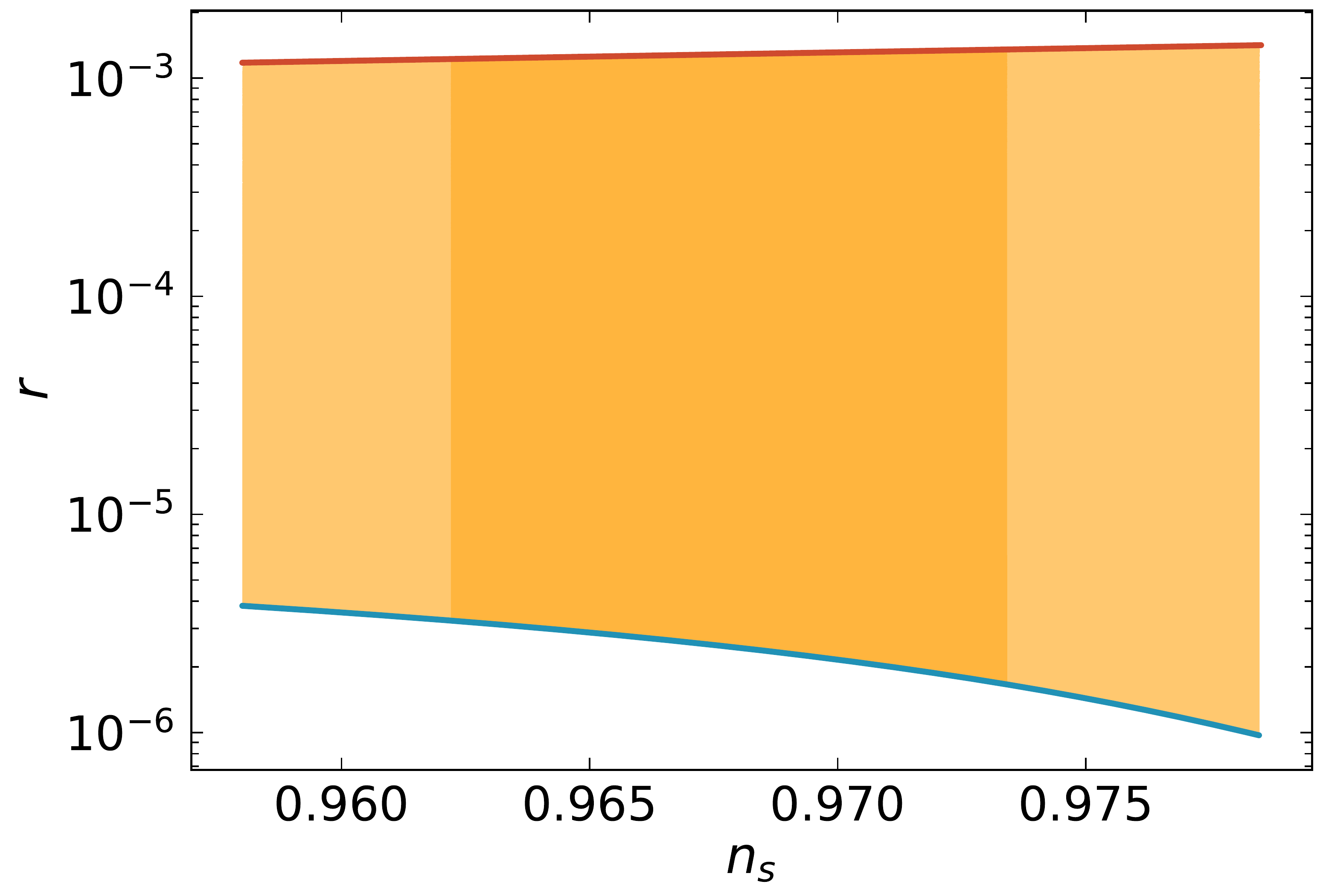}
	\centering \includegraphics[width=8.035cm]{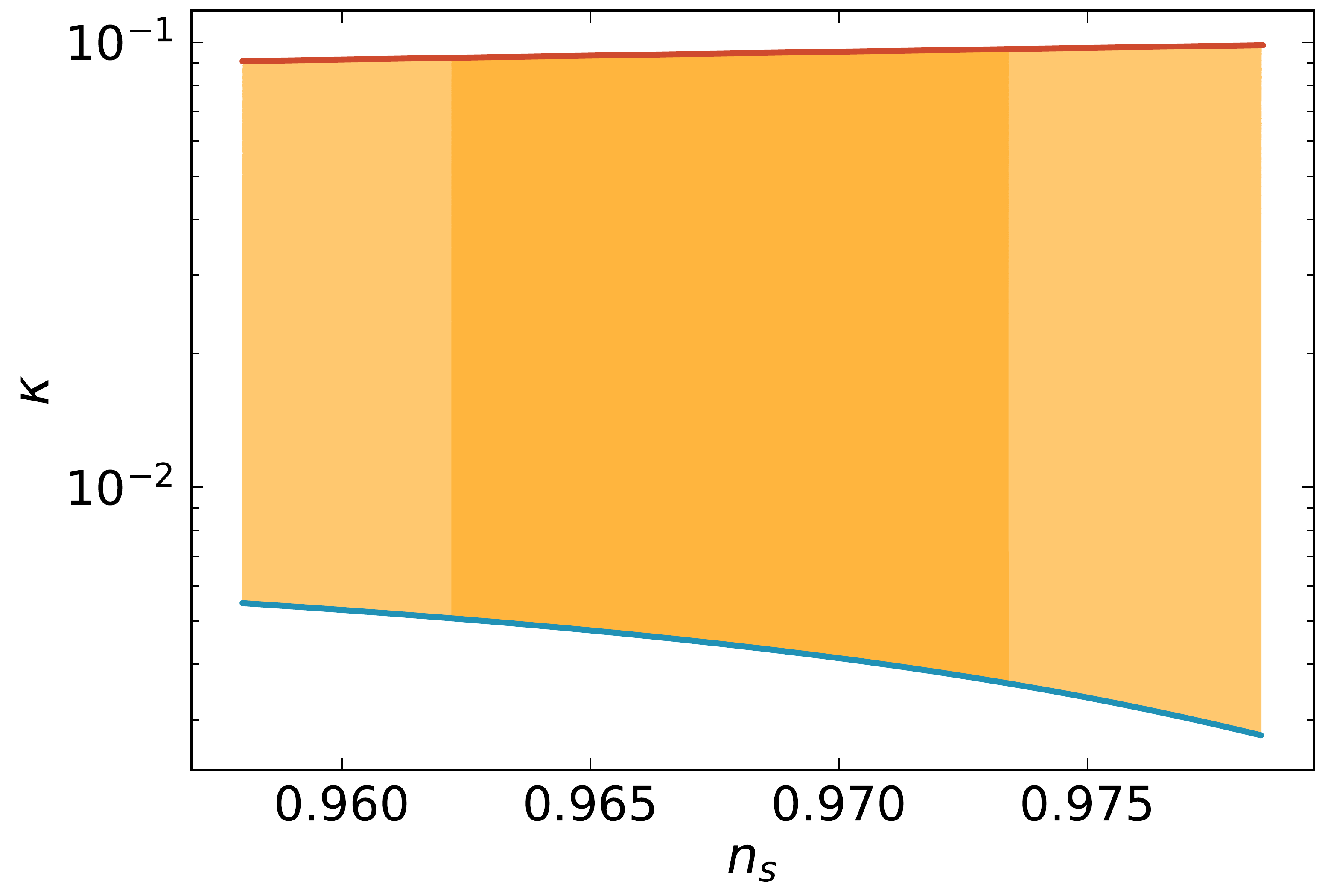}
	\caption{Behavior of $\kappa$ (right) and tensor-to-scalar ratio $r$ (left) with respect to scalar spectral index $n_s$ for $SU(5)$ breaking scale $M_{\alpha} \simeq M_{\text{GUT}} = 2 \times 10^{16}$ GeV. The lighter (darker) shaded region represents the Planck 2-$\sigma$ (1-$\sigma$) bounds, whereas the red and blue curves correspond to the $S_0 = m_P$ and $\kappa_{SS} = 1$ constraints, respectively.}
	\label{fig:nonminimal_1}
\end{figure}

\begin{figure}[!htb]
	\centering \includegraphics[width=8.035cm]{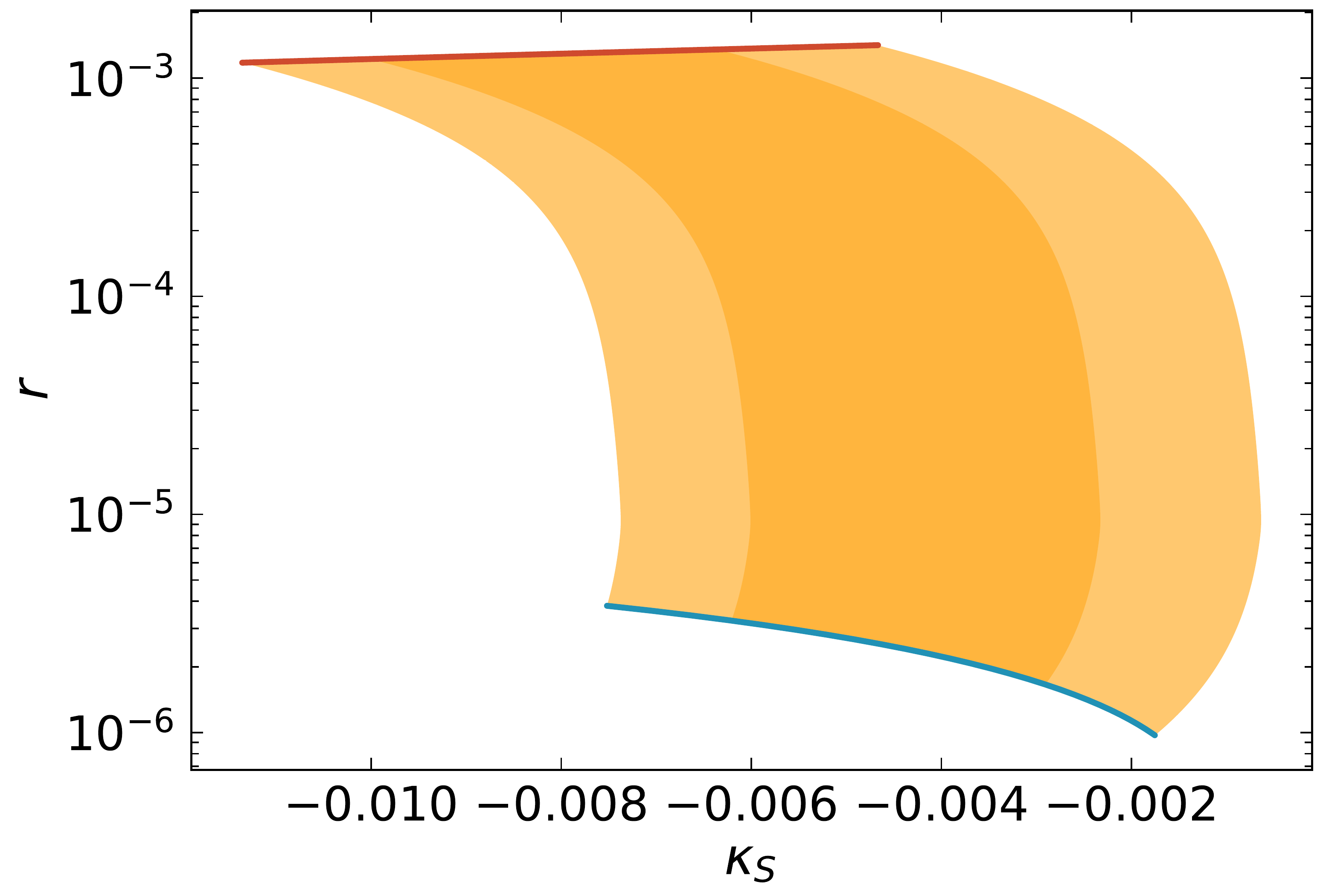}
	\centering \includegraphics[width=8.035cm]{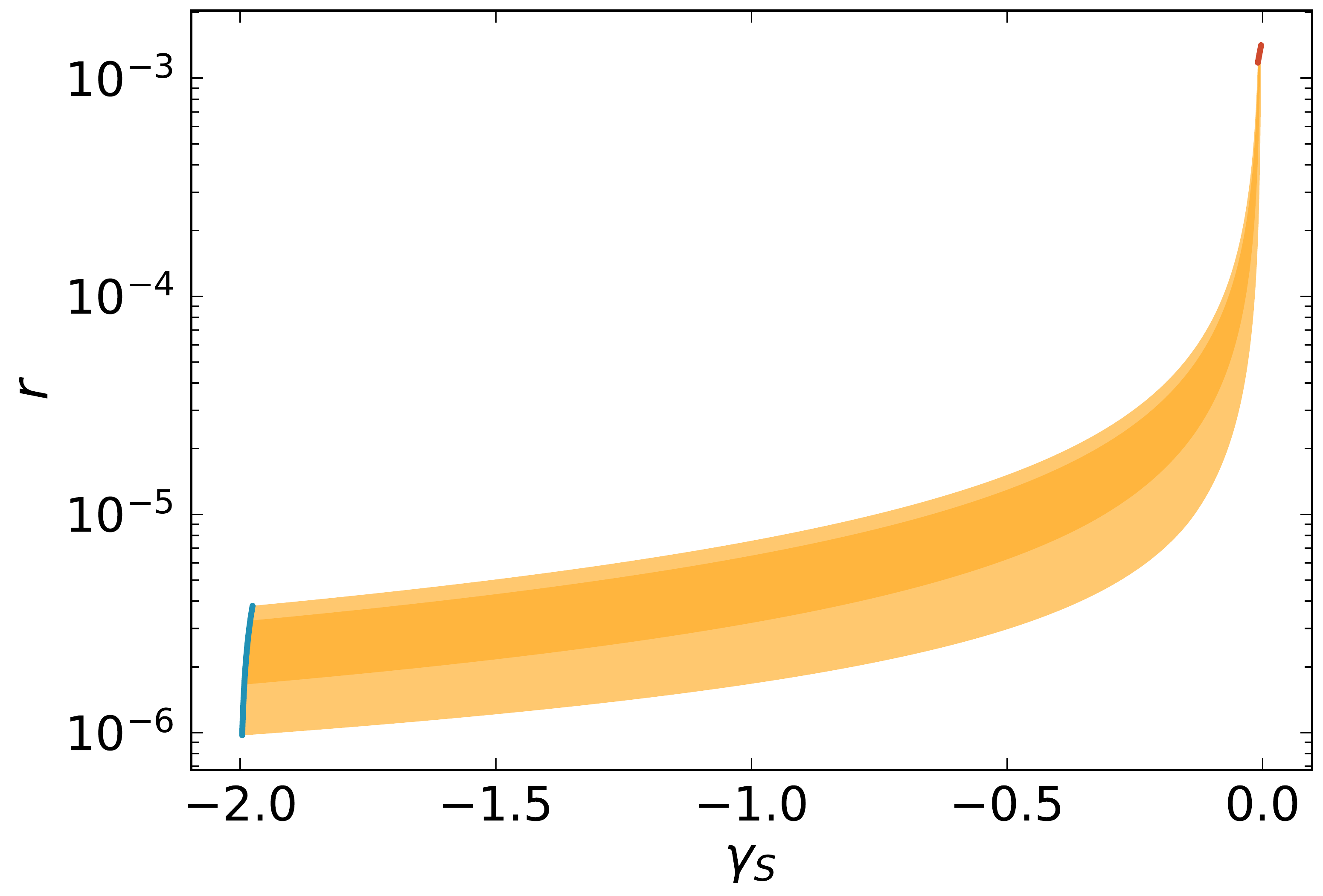}
	\caption{Behavior of tensor-to-scalar ratio $r$ with respect to the non-minimal coupling $\kappa_S$ (left) and quartic coupling $\gamma_S$ (right) for $SU(5)$ breaking scale $M_{\alpha} \simeq M_{\text{GUT}} = 2 \times 10^{16}$ GeV. The lighter (darker) shaded region represents the Planck 2-$\sigma$ (1-$\sigma$) bounds, whereas the red and blue curves correspond to the $S_0 = m_P$ and $\kappa_{SS} = 1$ constraints, respectively.}
	\label{fig:nonminimal_2}
\end{figure}

\begin{figure}[t]
	\centering \includegraphics[width=8.035cm]{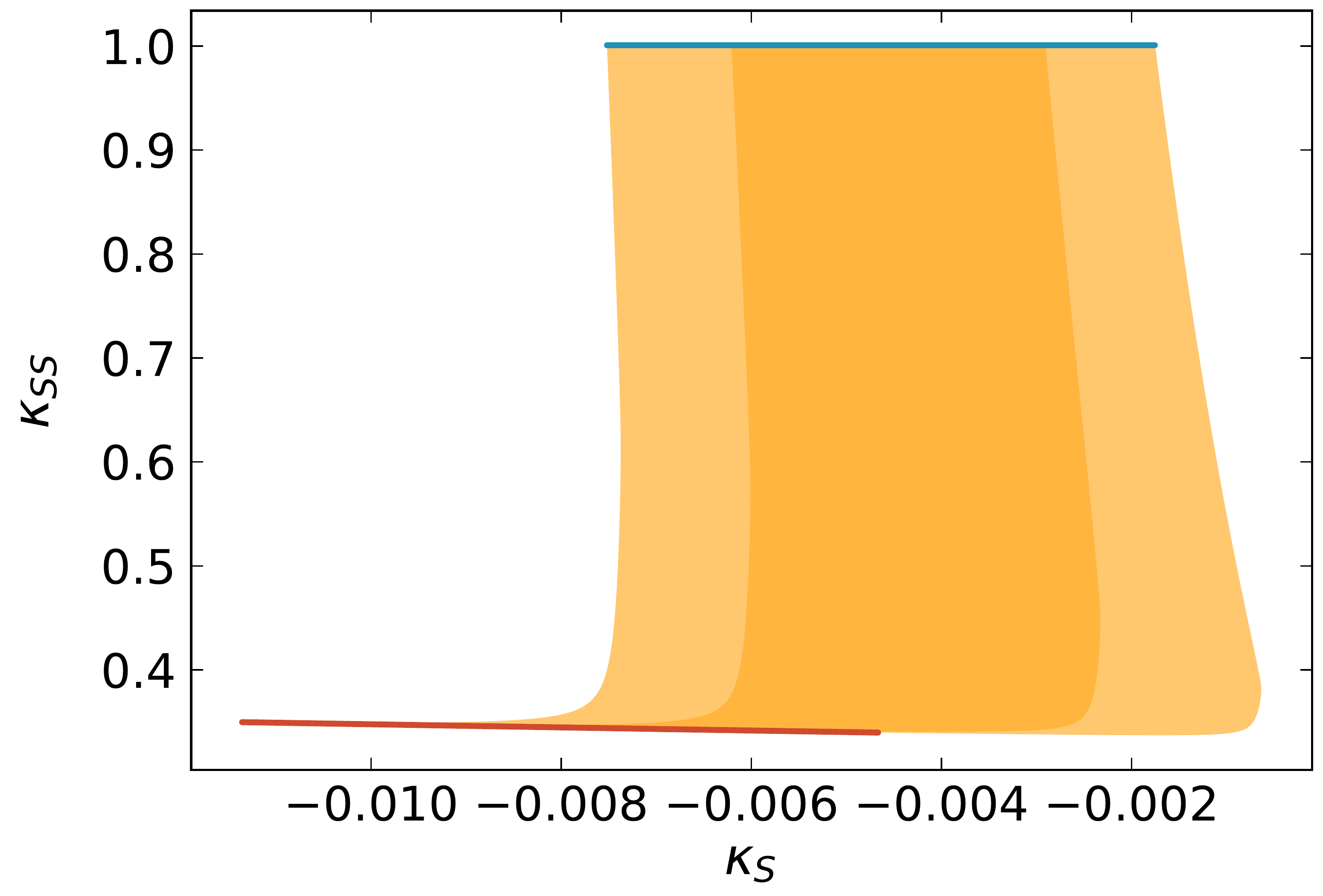}
	\centering \includegraphics[width=8.035cm]{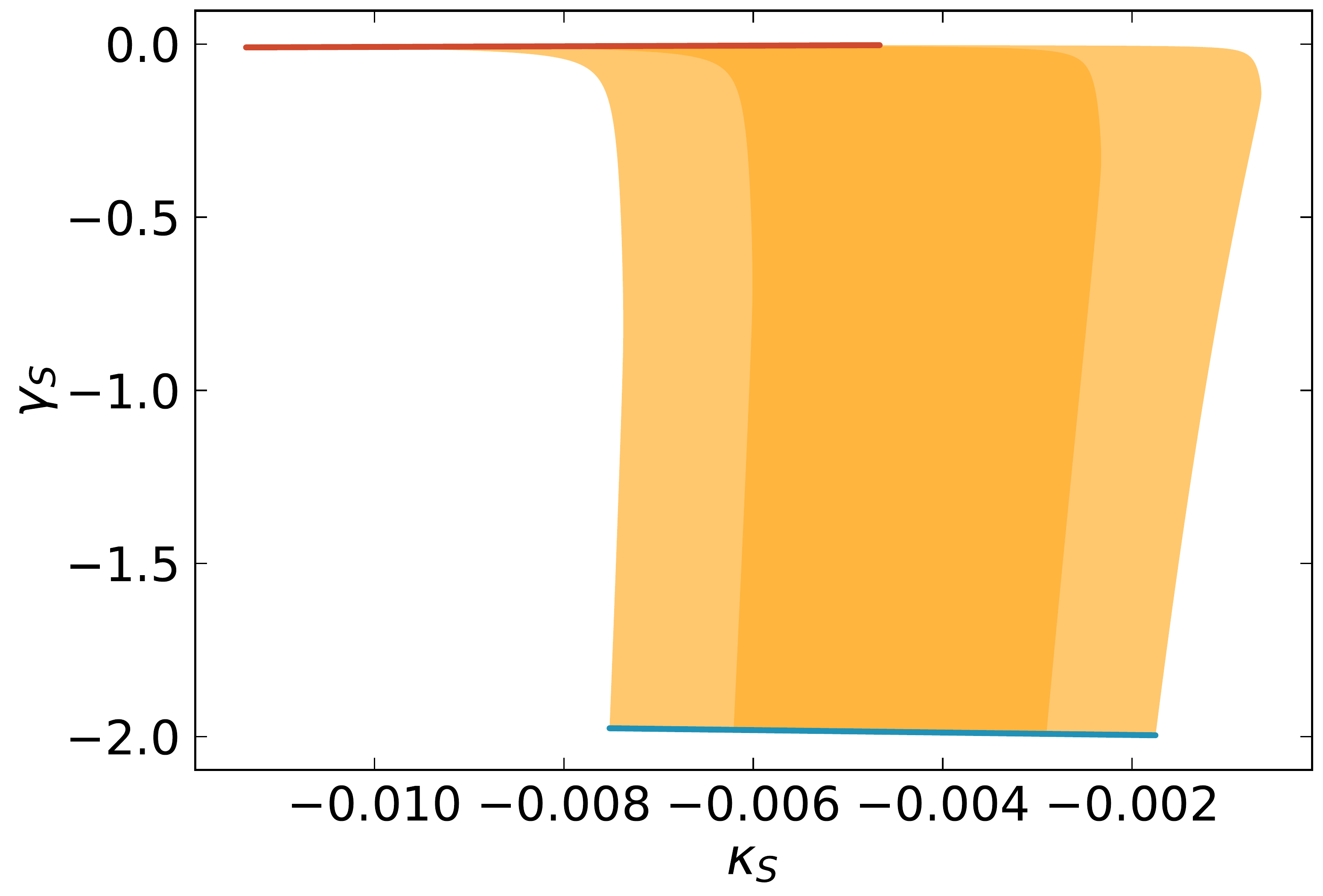}
	\caption{Behavior of non-minimal coupling $\kappa_{SS}$ (left) and quartic coupling $\gamma_S$ (right) with respect to the non-minimal coupling $\kappa_S$ for $SU(5)$ breaking scale $M_{\alpha} \simeq M_{\text{GUT}} = 2 \times 10^{16}$ GeV. The lighter (darker) shaded region represents the Planck 2-$\sigma$ (1-$\sigma$) bounds, whereas the red and blue curves correspond to the $S_0 = m_P$ and $\kappa_{SS} = 1$ constraints, respectively.}
	\label{fig:nonminimal_3}
\end{figure}

In this section we employ a non-minimal K\"ahler potential including non-renormalizable terms up to sixth order;

\begin{equation}
	\label{eq:nonminK}
	\begin{split} 
		K &= \vert S \vert^2 + \Tr \vert \Phi \vert^2 + \vert h \vert^2 + \vert \bar{h}\vert^2 + \vert \chi \vert^2 + \vert \bar{\chi} \vert^2
		\\ 
		& +\kappa_{S\Phi} \frac{\vert S\vert^2 \, \Tr \vert \Phi \vert^2}{m_P^2}
		+ \kappa_{S h} \frac{\vert S \vert^2 \vert h \vert^2}{m_P^2}
		+ \kappa_{S \bar{H}} \frac{\vert S \vert^2 \vert \bar{H} \vert^2}{m_P^2}
		+ \kappa_{S {\chi}} \frac{\vert S \vert^2 \vert \chi \vert^2}{m_P^2} + \kappa_{S \bar{\chi}} \frac{\vert S \vert^2 \vert \bar{\chi} \vert^2}{m_P^2}  
		\\ 
		& + \kappa_{H \Phi} \frac{\vert h \vert^2 \, \Tr \vert \Phi \vert^2}{m_P^2}
		+ \kappa_{h \chi} \frac{\vert h \vert^2  \vert \chi \vert^2}{m_P^2} + \kappa_{h \bar{\chi}} \frac{\vert h \vert^2  \vert \bar{\chi} \vert^2}{m_P^2} + \kappa_{\bar{h} \Phi} \frac{\vert \bar{h} \vert^2 \, \Tr \vert \Phi \vert^2}{m_P^2} 
		\\
		& + \kappa_{\bar{h} \chi} \frac{\vert \bar{h} \vert^2  \vert \chi \vert^2}{m_P^2} + \kappa_{\bar{h} \bar{\chi}} \frac{\vert \bar{h} \vert^2  \vert \bar{\chi} \vert^2}{m_P^2} + \kappa_{h \bar{h}} \frac{\vert h \vert^2 \vert \bar{h} \vert^2}{m_P^2}
		+ \kappa_{\chi \bar{\chi}} \frac{\vert \chi \vert^2 \vert \bar{\chi} \vert^2}{m_P^2} + \kappa_S \frac{\vert S\vert^4}{4 m_P^2} 
		\\
		& + \kappa_{\Phi} \frac{ (\Tr \vert \Phi \vert^2)^2}{4 m_P^2}
		+ \kappa_{H} \frac{ \vert h \vert^4}{4 m_P^2} + \kappa_{\bar{h}} \frac{ \vert \bar{h} \vert^4}{4 m_P^2}
		+ \kappa_{\chi} \frac{ \vert \chi \vert^4}{4 m_P^2}
		+ \kappa_{\bar{\chi}} \frac{ \vert \bar{\chi} \vert^4}{4 m_P^2}
		\\
		& + \kappa_{SS} \frac{\vert S\vert^6}{6 m_P^4} + \kappa_{\Phi \Phi} \frac{ (\Tr \vert \Phi \vert^2)^3}{6 m_P^4}
		+ \kappa_{h h} \frac{ \vert h \vert^6}{6 m_P^4}
		+ \kappa_{\bar{h} \bar{h}} \frac{ \vert \bar{h} \vert^6}{6 m_P^4}
		+ \kappa_{\chi \chi} \frac{ \vert \chi \vert^6}{6 m_P^4} + \kappa_{\bar{\chi} \bar{\chi}} \frac{ \vert \bar{\chi} \vert^6}{6 m_P^4}
		+ \cdots.
	\end{split}
\end{equation}
Including the one loop radiative corrections and soft SUSY mass terms, the full scalar potential during inflation then reads as,
\begin{eqnarray}
	V &\simeq& V_{\text{SUGRA}} + V_{\text{1-loop}} + V_{\text{Soft}} \nonumber \\ 
	&\simeq& \kappa ^2 M_{\alpha}^4 \Bigg[1+ \left(\frac{4(1-\kappa_{S\Phi})}{9\,(4/27 - \alpha^2)} -\kappa_S\,x^2 \right) \left( \frac{M_{\alpha}}{m_P}\right)^{2} \nonumber \\
	&+&   \left( \frac{4((1-2\kappa_{S\Phi})^2+1+\kappa_{\Phi})}{81\,(4/27 - \alpha^2)^2} \right. \nonumber \\
	&+& \left. \frac{4((1-\kappa_{S\Phi})^2-\kappa_{S}(1-2\kappa_{S\Phi}))x^2}{9\,(4/27 - \alpha^2)} 	+\frac{\gamma _{S}\,x^4}{2}\right) \left( \frac{M_{\alpha}}{m_P}\right)^{4} \nonumber  \\ 
	&+& \frac{\kappa^2}{16 \pi^2} \left[ F(M_{\alpha}^2,x^2) + 11\times 25\,F(5 M_{\alpha}^2,5\,x^2) \right] + \frac{\sigma_{\chi}^2}{8 \pi^2} F(M_{\alpha}^2,y^2) \nonumber \\ 
	&+&  \frac{a m_{3/2} x}{\kappa M_{\alpha}}  + \frac{M_S^2\, x^2}{\kappa^2 M_{\alpha}^2}  + \frac{8 M_{\phi}^2}{9 \kappa^2 M_{\alpha}^2 \left( 4/27 - \alpha^2 \right)} \Bigg] ,
\end{eqnarray}
\noindent where $\gamma_S = 1 - \frac{7 \kappa_S}{2} + 2 \kappa_S^2 - 3 \kappa_{SS}$. The results of our numerical calculations with a non-minimal K\"ahler potential are displayed in Figs. \ref{fig:nonminimal_1} - \ref{fig:nonminimal_3}. In obtaining these results, we have used up to second order approximation on the slow-roll parameters and the $SU(5)$ gauge symmetry breaking scale $M_{\alpha}$ is fixed at $M_{\text{GUT}} \simeq 2 \times 10^{16}$ GeV. We have also fixed the soft SUSY masses at $m_{3/2} \simeq M_S \simeq 10$ TeV, with $a = 1$ and $M_S^2 > 0$. 

As compared to the minimal case, the non-minimal K\"ahler potential increases the parametric space and with the addition of new parameters, we now expect to obtain $n_s$ within the latest Planck bounds with large values of tensor-to-scalar ratio $r$. The radiative corrections and SUGRA corrections parameterized by $\kappa_{S}$ and $\kappa_{SS}$, dominate the global SUSY potential while the soft mass terms with $m_{3/2} \simeq M_S \simeq 10$ TeV are adequately suppressed. To keep the SUGRA expansion under control we impose $S_0 \leq m_P$. We also restrict the non-minimal couplings $\vert \kappa_{S} \vert \leq 1$ and $\vert \kappa_{SS} \vert \leq 1$. These two constraints are shown in Figs. \ref{fig:nonminimal_1} - \ref{fig:nonminimal_3} by the red ($S_0 = m_P$) and blue ($\kappa_{SS} = 1$) curves. The lighter (darker) yellow region represents the Planck 2-$\sigma$ (1-$\sigma$) bounds on scalar spectral index $n_s$. By employing non-minimal K\"ahler potential, there is a significant increase in the tensor-to-scalar ratio $r$ and both $\kappa_S$ and $\gamma_{S}$ play vital role to bring the scalar spectral index $n_s$ within Planck 2-$\sigma$ data bounds, with a large value of tensor to scalar ratio $r \simeq 10^{-3}$. 

The behavior of tensor-to-scalar ratio $r$ and $\kappa$, as displayed in Fig. \ref{fig:nonminimal_1}, can be understood from the explicit relation \eqref{eq:rkappaexplicit} between $r$, $\kappa$ and $M_{\alpha}$ which shows that larger values of $r$ are expected when $\kappa$ or $M_{\alpha}$ is large. Since $M_{\alpha}$ is fixed, larger $r$ values should be obtained for large $\kappa$.  For fixed $M_{\alpha} \simeq 2 \times 10^{16}$ GeV, the largest value of $r$ ($\sim 1.5 \times 10^{-3}$) obtained in our numerical results occurs for $\kappa \simeq 0.1$. The behavior of tensor to scalar ratio $r$ with respect to $\kappa_{S}$ and $\gamma_{S}$ is presented in Fig. \ref{fig:nonminimal_2}, while Fig. \ref{fig:nonminimal_3} depicts the behavior of $\kappa_{SS}$ and $\gamma_{S}$ with respect to $\kappa_{S}$. It can be seen that the large $r$ values are obtained with non-minimal couplings $\kappa_S < 0$, $\kappa_{SS} > 0$ and the quartic coupling $\gamma_S <0 $. Moreover, in the large $r$ limit, both $\kappa_{S}$ and $\kappa_{SS}$ are tuned to make $\gamma_{S}$ very small ($\sim -0.003$). Note that large tensor modes can be obtained for any value of scalar spectral index $n_s$ within Planck 2-$\sigma$ bounds. Finally, smaller $r$ values ($\sim 10^{-6}$) are obtained for $S_0 \lesssim 0.05 \, m_P$ and $\kappa_{SS} \simeq 1$ for which $\gamma_{S}$ is negative and fairly large ($\sim -2$).

The spectral index $n_s$ and tensor to scalar ratio $r$ in the leading order slow-roll approximation are given by
\begin{equation}
	n_s \simeq 1 - 2 \kappa_S + \left( 6 \gamma_S x_0^2 + \frac{8 \left(1 - \kappa_S \right)}{9 \left( 4/27 - \alpha^2  \right)} \right) \left(\frac{M_{\alpha}}{m_P}\right)^2 + \frac{278 \kappa^2 F^{''}(5 x_0)}{16 \pi^2} \left(\frac{m_P}{M_{\alpha}}\right)^2 , 
\end{equation}
\begin{equation}
	r \simeq 4 \left( \frac{m_P}{M_{\alpha}} \right)^2 \left( - 2 \kappa_S x_0 \left(\frac{M_{\alpha}}{m_P}\right)^2  + \left( 2 \gamma_S x_0^3 + \frac{8 \left(1 - \kappa_S \right)}{9 \left( 4/27 - \alpha^2  \right)} \right) \left(\frac{M_{\alpha}}{m_P}\right)^4 + \frac{278 \kappa^2 F^{'}(5 x_0)}{16 \pi^2}  \right)^2.
\end{equation}
Solving these two equations simultaneously for $S_0 \simeq m_P$, $r \simeq 10^{-3}$, and $n_s \simeq 0.9655$ we obtain $\kappa_S \simeq -0.006$ and $\gamma_{S} \simeq -0.005$. Similarly in the small $r$ region for $S_0 \simeq (0.05)  m_P$, $r \simeq 3 \times 10^{-6}$, and $n_s \simeq 0.9655$ we obtain $\kappa_S \simeq -0.005$ and $\gamma_{S} \simeq -2$. These approximate
values are very close to the actual values obtained in the numerical calculations. The above analytical equations therefore gives a valid approximation of our numerical results displayed in Figs. \ref{fig:nonminimal_1} - \ref{fig:nonminimal_3}. For non-minimal couplings ($-0.011 \lesssim \kappa_S \lesssim - 0.00063$) and ($0.34 \lesssim \kappa_{SS} \lesssim 1$), we obtain the scalar spectral index $n_s$ within the Planck 2-$\sigma$ bounds and tensor to scalar ratio $r$ in the range ($9.7 \times 10^{-7} \lesssim r \lesssim 1.5 \times 10^{-3}$).

\section{\large{\bf Radiative Breaking of $U(1)_{\chi}$} Symmetry}\label{sec7}

After the end of inflation, the effective unbroken gauge symmetry is $SU(3)_C \times SU(2)_L \times U(1)_Y \times U(1)_{\chi}$. The charge assignments of the fields under this symmetry are displayed in Table \ref{tab:table_radiative}.
\begin{table}[!htb]
	\setlength\extrarowheight{3pt}
	\centering
	\begin{tabular}{c c}
		\hline \hline \rowcolor{Gray}
		\multicolumn{1}{c}{}                    & \multicolumn{1}{c}{}                    \\ \rowcolor{Gray}
	\multicolumn{1}{c}{\multirow{-2}{*}{Superfields}} & \multirow{-2}{*}{\begin{tabular}[c]{@{}c@{}}Representations under\\ $SU(3)_C \times SU(2)_L \times U(1)_Y \times U(1)_{\chi}$\end{tabular}} \\ 
	\hline
	\rowcolor{Gray2} \multicolumn{2}{c}{Matter sector}                                      \\ \hline
	\multicolumn{1}{c}{$Q$}     &       $\left( \mathbf{3}, \mathbf{2}, 1/6, -1 \right)$          \\
	\multicolumn{1}{c}{$u^c$} &  $\left( \bar{\mathbf{3}}, \mathbf{1}, -2/3, -1 \right)$\\
	\multicolumn{1}{c}{$d^c$}	&   $\left( \bar{\mathbf{3}}, \mathbf{1}, 1/3, 3 \right)$ \\
	\multicolumn{1}{c}{$\ell$}	&   $\left( \mathbf{1}, \mathbf{2}, -1/2, 3 \right)$ \\
	\multicolumn{1}{c}{$e^c$}	&   $\left( \mathbf{1}, \mathbf{1}, 1, -1 \right)$ \\
	\multicolumn{1}{c}{$\nu^c$}	&   $\left( \mathbf{1}, \mathbf{1}, 0, -5 \right)$  \\ \hline
	\rowcolor{Gray2} \multicolumn{2}{c}{Scalar sector}                               \\ \hline
	\multicolumn{1}{c}{$h_u$}	&   $\left( \mathbf{1}, \mathbf{2}, 1/2, 2 \right)$ \\
	\multicolumn{1}{c}{$h_d$}	&   $\left( \mathbf{1}, \mathbf{2}, -1/2, -2 \right)$ \\
	\multicolumn{1}{c}{$\chi$}	&   $\left( \mathbf{1}, \mathbf{1}, 0, 10 \right)$ \\
	\multicolumn{1}{c}{$\bar{\chi}$} & $\left( \mathbf{1}, \mathbf{1}, 0, -10 \right)$ \\
	\multicolumn{1}{c}{$S$}	& $\left( \mathbf{1}, \mathbf{1}, 0, 0 \right)$ \\  \hline \hline
	\end{tabular}
	\caption{Superfields and their representations under the effective unbroken guage symmetry $SU(3)_C \times SU(2)_L \times U(1)_Y \times U(1)_{\chi}$ after the end of inflation.}
	\label{tab:table_radiative}
\end{table}
The superpotential terms relevant for $U(1)_{\chi}$ symmetry breaking are given by

\begin{eqnarray}
	\label{W_u1_breaking} \nonumber
	W &=& W_{\text{MSSM}} + W_{\chi} \\
W_{\chi} &=& \sigma_{\chi} S \chi \bar{\chi} 
+ \lambda_{ij} \chi \nu_{i}^c \nu_{j}^c \,.
\end{eqnarray}
From the above equation, we obtain
\begin{eqnarray} \nonumber
	F_S^{\dagger} &=& \frac{\partial W}{\partial S} = \sigma_{\chi} \chi \bar{\chi} = 0 ,\\ \nonumber
	F_{\bar{\chi}}^{\dagger} &=& \frac{\partial W}{\partial \bar{\chi}} = \sigma_{\chi} S \chi  = 0 ,\\ \nonumber
	F_{\chi}^{\dagger} &=& \frac{\partial W}{\partial \chi} = \sigma_{\chi} S \bar{\chi} + \lambda_{ij} \nu_{i}^c \nu_{j}^c = 0 ,\\
	F_{\nu^c}^{\dagger} &=& \frac{\partial W}{\partial \nu^c} = 2 \lambda_{ij} \nu_{i}^c \nu_{j}^c = 0 .
\end{eqnarray}
This leads to the following vacua;
\begin{equation}
	\langle \chi \rangle = \langle \bar{\chi} \rangle= 0, \qquad \langle \nu_i^c \rangle = 0 , \qquad \langle S \rangle = \text{Arbitrary}.
\end{equation}
In order to break the $U(1)_{\chi}$ symmetry, a non-zero VEV of the field $\chi$ is desired; $\langle \chi \rangle = \langle \bar{\chi} \rangle \neq 0$.  Including the soft SUSY breaking mass terms,
\begin{eqnarray} \nonumber
	V_{\text{Soft}} &=& m_S^2 \vert S \vert^2 + m_{\chi}^2 \vert \chi \vert^2 + m_{\bar{\chi}}^2 \vert \bar{\chi} \vert^2 + m_{\nu_i}^2 \vert \nu_i^c \vert^2 \\
	&+& A_{\nu} \lambda_{ij} \chi \nu_{i}^c \nu_{j}^c + A_{l} y_{ij}^{\nu} \nu_i^c l_j h_u + A_{\chi} \sigma_{\chi} S \chi \bar{\chi} + \frac{1}{2} M_{\chi} Z_{\chi} Z_{\chi} , 
\end{eqnarray}
where $A_{\nu}$ and $A_{\chi}$ are coefficients of linear soft mass terms, $Z_{\chi}$ is the $U(1)_{\chi}$ gaugino and $M_{\chi}$ is the gaugino mass. The full scalar potential is then given by,
\begin{eqnarray}
	\nonumber
	V &=& V_{F} + V_{D} + V_{\text{Soft}} \\ \nonumber
	&=&  \sigma_{\chi}^2 \vert \chi \bar{\chi} \vert^2 +  \left| \, \sigma_{\chi} S \bar{\chi} + \lambda_{ij}  \nu_{i}^{c}\nu_{j}^{c}\, \right|^2 + \left| \, \sigma_{\chi} S \chi \, \right|^2 + \left|2 \lambda_{ij} {\chi} \nu_{i}^{c}\right|^2 \\ \nonumber
	&+& 50 g_{\chi}^2 \left( \vert \chi \vert^2 - \vert \bar{\chi} \vert^2 \right)^2 \\ \nonumber
	 &+& m_S^2 \vert S \vert^2 + m_{\chi}^2 \vert \chi \vert^2 + m_{\bar{\chi}}^2 \vert \bar{\chi} \vert^2 + m_{\nu_i}^2 \vert \nu_i^c \vert^2 \\
	&+& A_{\nu} \lambda_{ij} \chi \nu_{i}^c \nu_{j}^c + A_{\chi} \sigma_{\chi} S \chi \bar{\chi} + \frac{1}{2} M_{\chi} Z_{\chi} Z_{\chi}.
\end{eqnarray}
The potential minima can be obtained as follows;
\begin{eqnarray} \nonumber
 \frac{\partial V}{\partial S^{\dagger}} &=& \sigma_{\chi}^2 S\left( \vert \chi \vert^2+\vert \bar{\chi} \vert^2\right)+\sigma_{\chi} \lambda_{ij}  \nu_{i}^{c}\nu_{j}^{c}\bar{\chi}^{\dagger} + m_{S}^2 S = 0 ,\\ \nonumber
\frac{\partial V}{\partial \bar{\chi}^{\dagger}}&=&\sigma_{\chi}^2\left( \chi\vert \chi \vert^2 +\bar{\chi} \vert S \vert^2\right)-100 g_{\chi}^2 \bar{\chi}\left( \vert \chi \vert^2 - \vert \bar{\chi} \vert^2 \right)+\sigma_{\chi} \lambda_{ij}  \nu_{i}^{c}\nu_{j}^{c}S^{\dagger} + m_{\bar{\chi}}^2  \bar{\chi}= 0 ,\\ \nonumber
\frac{\partial V}{\partial \chi^{\dagger}}&=& \sigma_{\chi}^2\left( \bar{\chi}\vert \chi \vert^2 +\chi \vert S \vert^2\right)+100 g_{\chi}^2 \chi \left( \vert \chi \vert^2 - \vert \bar{\chi} \vert^2 \right)+4  \lambda_{ij}^2  \vert \nu_i^c \vert^2\chi + m_{\chi}^2  \chi= 0,\\
 \frac{\partial V}{\partial {\nu_i^c}^{\dagger}} &=& 2{\nu_i^c}^{\dagger}\left(\sigma_{\chi} S \bar{\chi} +\lambda_{ij} \nu_{i}^c \nu_{j}^c\right) +4  \lambda_{ij}^2  \nu_{i}^{c} \vert \chi \vert^2 + m_{\nu_i}^2 \nu_i^c = 0 .
\end{eqnarray}

 Conservation of $R$-parity requires, $\langle \nu_i ^c \rangle = 0$. The VEV of the fields $\chi$, $\bar{\chi}$ is found to be,
\begin{equation}
	\langle \vert \bar{\chi} \vert\rangle=\langle \vert \chi \vert\rangle = \sqrt{-\frac{m_S^2}{2 \, \sigma_{\chi}^2}}\;. 
\end{equation}
The negative mass squared, $m_{S}^2 < 0$ should be satisfied at an intermediate scale $M_{*}$ below the GUT scale to realize the correct $U(1)_{\chi}$ symmetry breaking. A negative mass squared can be achieved through the RG running from the GUT scale to an intermediate scale with a large enough Yukawa coupling even if the mass squared is positive at the GUT scale. 
We consider the $U(1)_{\chi}$ renormalization group equations and analyze the running of the scalar masses $m_{\chi}^2$, $m_{\nu_i^c}^2$, $m_{\bar{\chi}}^2$ and $m_{S}^2$. A negative mass-squared $m_{S}^2$ will trigger the radiative breaking of $U(1)_{\chi}$ symmetry. We show that the mass-squared of the fields $\chi$, $\bar{\chi}$, $\nu^c$ and $S$ evolve in such a way that $m_S^2$ becomes negative whereas $m_{\nu_i^c}^2$, $m_{\chi}^2$ and $m_{\bar{\chi}}^2$ remain positive.
The renormalization group equations are given by
\begin{eqnarray} 
	16 \pi^2 \frac{d g_{\chi}}{d t} &=& \frac{57}{5} g_{\chi}^3 , \\
	16 \pi^2 \frac{d M_{\chi}}{d t} &=& \frac{114}{5} g_{\chi}^2 M_{\chi} , \\
	16 \pi^2 \frac{d \lambda_{i}}{d t} &=& \lambda_{i} \left( 8 \lambda_{i}^2 + 2 \Tr \lambda^2 + \sigma_{\chi}^2 - \frac{15}{2} g_{\chi}^2  \right) , \\
	16 \pi^2 \frac{d \sigma_{\chi}}{d t} &=& \sigma_{\chi} \left( 3 \sigma_{\chi}^2 + 2 \Tr \lambda^2 - 10 g_{\chi}^2  \right) , \\
	16 \pi^2 \frac{d m_{\chi}^2}{d t} &=&  2 \sigma_{\chi}^2 \left( m_{\chi}^2 + m_{\bar{\chi}}^2 + m_S^2  \right) + 4 m_{\chi}^2 \Tr \lambda^2 + 8  \Tr \left(m_{\nu^c}^2\lambda^2\right) \\ &+& 4 T_{\sigma_{\chi}}^2  - 20 g_{\chi}^2 M_{\chi}^2 ,  \\ 
	16 \pi^2 \frac{d m_{\bar{\chi}}^2}{d t} &=& 2 \sigma_{\chi}^2 \left( m_{\chi}^2 + m_{\bar{\chi}}^2 + m_S^2  \right) + 2 T_{\sigma_{\chi}}^2  - 20 g_{\chi}^2 M_{\chi}^2  , \\ 
	16 \pi^2 \frac{d m_{\nu^c_{i}}^2}{d t} &=& 8 \lambda_i^2 \left( m_{\chi}^2 + 2 m_{\nu^c_{i}}^2 \right) + 8 T_{\nu i}^2 - 5 g_{\chi}^2 M_{\chi}^2 , \\
	16 \pi^2 \frac{d m_{S}^2}{d t} &=& 2 \sigma_{\chi}^2 \left( m_{\chi}^2 + m_{\bar{\chi}}^2 + m_S^2  \right) + 2 T_{\sigma_{\chi}}^2 , \\
	16 \pi^2 \frac{d T_{\sigma_{\chi}}}{d t} &=& T_{\sigma_{\chi}} \left( 9 \sigma_{\chi}^2 + 2 \Tr \lambda^2 - 10 g_{\chi}^2 \right) + 4 \sigma_{\chi} \left( \Tr \lambda^2 + 5 g_{\chi}^2 M_{\chi}\right) ,\\
	16 \pi^2 \frac{d T_{\nu_i}}{d t} &=& T_{\nu_i} \left( \sigma_{\chi}^2 + 2 \Tr \lambda^2 + 24 \lambda_i^2 - \frac{15}{2} g_{\chi}^2  \right) \\ &+& 2 \lambda_i \left( 2 \Tr \lambda^2 + 15 g_{\chi}^2 M_{\chi} +  T_{\sigma_{\chi}} \sigma_{\chi} + 2 \Tr (\lambda_i T_{\nu_i}) \right),
\end{eqnarray}
\begin{figure}[!htb]
	\centering \includegraphics[width=10.0cm]{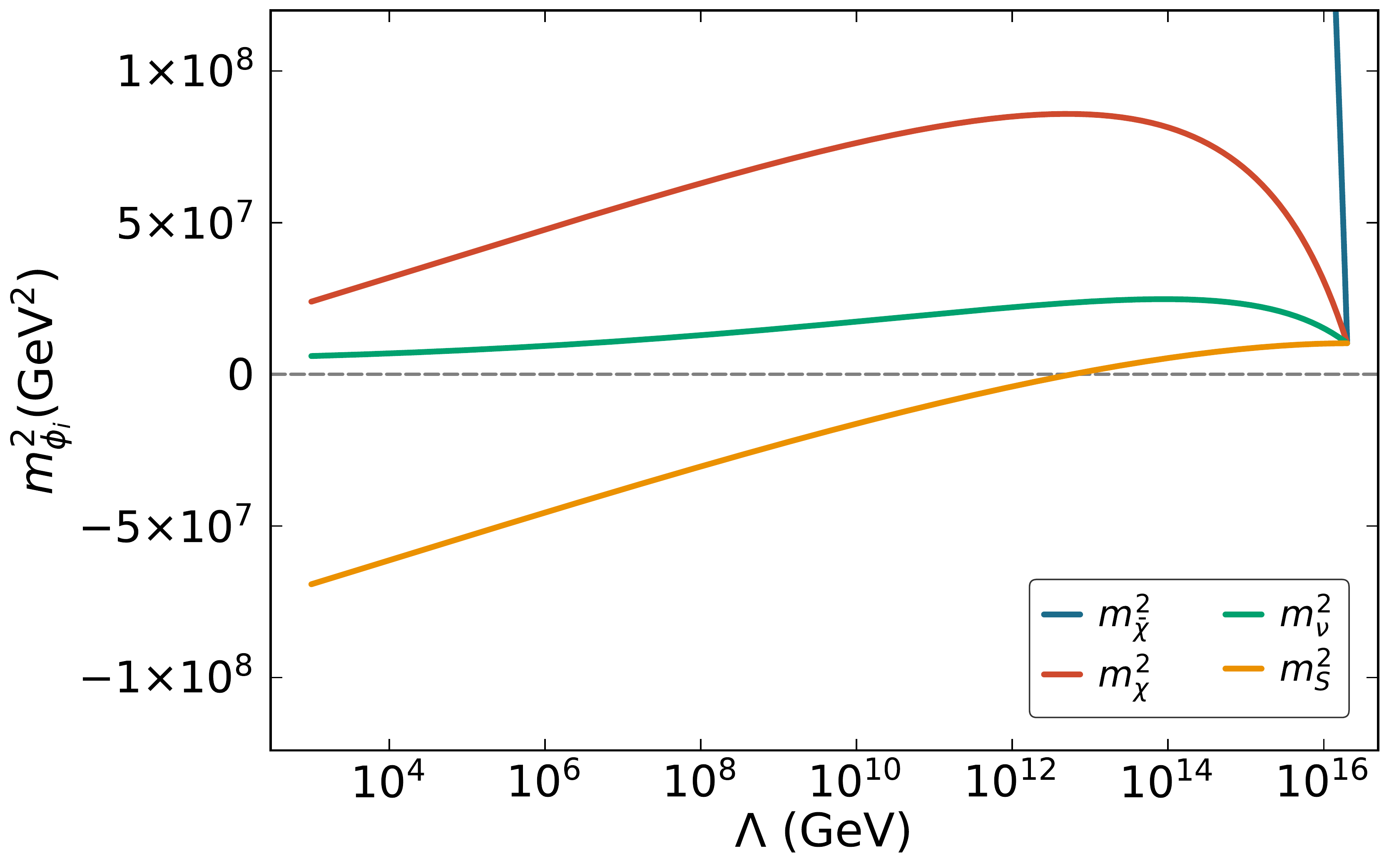}
	\caption{The evolution of scalar squared masses $m_{\chi}^2$, $m_{\nu^c}^2$, $m_{\bar{\chi}}^2$ and $m_{S}^2$ from the GUT to TeV scale.}
	\label{RGEs}
\end{figure}
where $\lambda_{ij} = \diag \left( \lambda_1, \lambda_2, \lambda_3\right)$. The evolution of these parameters depends on the boundary conditions at GUT scale, $M_{\text{GUT}}=2\times 10^{16}$ GeV. We assume universal soft SUSY breaking at this scale,
\begin{gather} \label{mzb}
	\quad m_{\chi}^2 = m_{\bar{\chi}}^2 = m_{\nu_i^c}^2 = m_{S}^2 = m_0^2, \quad M_{\chi} =16 ~ \text{TeV}, \nonumber \\ 
	g_{\chi}^2=1.04,\quad \lambda_{3} = 0.4, \quad \sigma_{\chi} = 0.3, \quad m_0 = 3.2 ~ \text{TeV}, \\ \nonumber
	T_{\nu_i} = \lambda_{i} A_{\nu}, \quad T_{\sigma_{\chi}} = \sigma_{\chi} A_{\chi} .
\end{gather}
Here we let the tri-linear couplings ($A_{\nu} = A_{\chi} = 0$) vanish as they have negligible effect on the overall running of the parameters. Also, for simplicity, we neglect the couplings of the first two right handed neutrino (RHNs) generations ($\lambda_1 = \lambda_2 = 0$). Fig. \ref{RGEs} shows the running of scalar masses from GUT scale. It can be seen that the mass-squared $m_{S}^2$ turns negative at scale $\sim 8 \times 10^{12}$ GeV, whereas $m_{\bar{\chi}}^2$ rapidly increase and approaches $\sim 10^9 ~ \text{GeV}^2$ around TeV scale. Note that, although the running of mass squared $m_{\nu_i^c}^2$ and $m_{\chi}^2$ decreases from $M_{\text{GUT}}$, they always remain positive.

The breaking of $U(1)_{\chi}$ at the end of inflation yields topologically stable cosmic strings on which, the observational bounds are given in terms of the dimensionless quantity $G_N \mu_s$, which characterizes the strength of the gravitational interaction of the strings. Where $G_N$ is the Newton’s constant, and $\mu_s \simeq 2 \pi \langle \chi \rangle^2$ denotes the mass per unit length of the string. The Planck bound on $G_N \mu_s$ derived from constraints on the string contribution to the CMB power spectrum is given by \cite{Ade:2013xla,Ade:2015xua}
\begin{equation}
	G_N \mu_s \lesssim 2.4 \times 10^{-7}.
\end{equation}
This then translates to the following upper bound on $U(1)_{\chi}$ breaking scale
\begin{equation}
	\langle \chi \rangle \lesssim 2.35 \times 10^{15} ~ \text{GeV},
\end{equation}
which is easily satisfied as can be seen from Fig. \ref{RGEs} and depends on the initial boundary conditions at GUT scale. After the $U(1)_{\chi}$ symmetry breaking, the RHNs and the $U(1)_{\chi}$ gauge boson $(Z^{\prime})$ acquire the following masses;
\begin{equation}
	M_{Z^{\prime}}^2\simeq g_{\chi}^2\langle \chi \rangle^2, \quad m_{\nu_3}^2\simeq \lambda_{3}^2 \langle \chi \rangle^2 ,
\end{equation}
which for the particular boundary condition in \eqref{mzb} yields, $M_{Z^{\prime}}\simeq 4.89$ TeV and $m_{\nu_3} \simeq 2.44$ TeV. The bound on $M_{Z^{\prime}}$ is constantly being updated by comprehensive analyses. The severest bound on $M_{Z'}$ comes from negative results of LEP data, $M_{Z'}/g_{\chi} \geq 6$ TeV \cite{Cacciapaglia:2006pk}. Even though considering its decay modes can lower the mass bound on $Z^{\prime}$ \cite{Accomando:2013sfa}, setting $M_{Z^{\prime}} \geq 4$ TeV \cite{ATLAS:2017wce} guarantees avoiding possible exclusions limit due to the light $Z^{\prime}$ mass.

The mass of RHNs predicted by the above model is of the order of TeV scale. Since they are singlet under the SM gauge group, a mixing between the RHNs and the SM neutrinos is generated through the Dirac Yukawa coupling in the seesaw mechanism. As a result, the RHN mass eigenstates  couple to the weak gauge bosons $W,Z$ through this mixing. 
Although in general, this mixing can be made sizable even for TeV-scale RHNs, contrary to the naive seesaw expectations, under special textures of the Dirac and RHN Majorana mass matrices~\cite{Pilaftsis:1991ug, Tommasini:1995ii, Gluza:2002vs, Xing:2009in, Gavela:2009cd, He:2009ua, Adhikari:2010yt, Deppisch:2010fr, Mitra:2011qr, Dev:2013oxa, Chattopadhyay:2017zvs},  it has been shown \cite{Das:2017nvm} that 
this mixing has an upper bound of ${\cal O}(0.01)$ to satisfy various experimental constraints, such as the neutrino oscillation data, the electroweak precision measurements, neutrinoless double beta decay 
and the charged lepton flavor violating (LFV) processes. 
Hence, the canonical production cross section of TeV-scale RHNs through either the weak gauge bosons~\cite{Datta:1993nm, Panella:2001wq, Han:2006ip, delAguila:2007qnc, Dev:2013wba, Alva:2014gxa, Das:2015toa, Das:2016hof, Pascoli:2018heg} or the Higgs boson~\cite{Dev:2012zg, Cely:2012bz, Hessler:2014ssa, Das:2017zjc, Das:2017rsu}  at the LHC is expected to be very small within the minimal seesaw. 

In the above model, all SM fermions as well as the RHNs have non-zero $U(1)_\chi$ charges, and therefore, the RHNs can be efficiently produced at colliders, in particular, through the resonant production of $Z^\prime$ boson, if kinematically allowed, and its subsequent decay into a pair of RHNs.

\section{\large{\bf Summary}}\label{sec8}
We have explored shifted hybrid inflation in the framework of supersymmetric $SU(5) \times U(1)_{\chi}$ model where $SU(5)$ gauge symmetry is spontaneously broken during inflation, inflating the disastrous magnetic monopoles away. The $U(1)_{\chi}$ symmetry is radiatevely broken after the end of inflation at an intermediate scale, yielding topologically stable cosmic strings. The symmetry breaking scale of $U(1)_{\chi}$ depends on the initial boundary conditions at the GUT scale and easily satisfies Planck's bound on $G_N \mu_s$. The $d = 5$ proton lifetime for the decay $p \rightarrow K^+ \bar{\nu}$, mediated by color-triplet Higgsinos is found to satisfy Super-Kamiokandae experimental bounds for SUSY breaking scale $M_{\text{SUSY}} \gtrsim 10$ TeV. We have shown that with minimal K\"ahler potential, the soft supersymmetry breaking terms play a vital role in bringing the scalar spectral index $n_s$ within the Planck's latest bounds and the $SU(5)$ guage symmetry breaking scale is obtained in the range ($2 \times 10^{15} \lesssim M_{\alpha} \lesssim 2 \times 10^{16}$) GeV with small values of tensor-to-scalar ratio $r \lesssim 2.5 \times 10^{-6}$. In a non-minimal K\"ahler potential setup, large values of tensor to scalar ratio are obtained ($r \lesssim 10^{-3}$) with non-minimal couplings ($-0.011 \lesssim \kappa_S \lesssim - 0.00063$) and ($0.34 \lesssim \kappa_{SS} \lesssim 1$) and symmetry breaking scale $M_{\alpha} \simeq M_{\text{GUT}} = 2 \times 10^{16}$ GeV.

\section*{Acknowledgements}
The authors would like to thank Mansoor Ur Rehman, Lorenzo Callibi, Shabbar Raza and Qaisar Shafi for helpful discussions. 



\begin{thebibliography}{99}


\bibitem{Dvali:1994ms}
G.~R.~Dvali, Q.~Shafi and R.~K.~Schaefer,
Phys.\ Rev.\ Lett.\  {\bf 73}, 1886 (1994)
[arXiv:hep-ph/9406319].


\bibitem{Copeland:1994vg}
E.~J.~Copeland, A.~R.~Liddle, D.~H.~Lyth, E.~D.~Stewart and D.~Wands,
Phys.\ Rev.\  D {\bf 49}, 6410 (1994)
[arXiv:astro-ph/9401011].


\bibitem{Linde:1997sj}
A.~D.~Linde and A.~Riotto,
Phys.\ Rev.\  D {\bf 56}, R1841 (1997)
[arXiv:hep-ph/9703209].

\bibitem{Linde:1993cn}
A.~D. Linde,
Phys. Rev. {\bf D49}, 748 (1994), arXiv:astro-ph/9307002.

\bibitem{Dvali:1997uq}
G.~R. Dvali, G.~Lazarides, and Q.~Shafi,
Phys. Lett. {\bf B424}, 259 (1998), arXiv:hep-ph/9710314.

\bibitem{Georgi:1974sy}
H.~Georgi and S.~L.~Glashow,
Phys. Rev. Lett. \textbf{32} (1974), 438-441

\bibitem{Barr:1981qv}
S.~M.~Barr,
Phys. Lett. B \textbf{112} (1982), 219-222

\bibitem{Antoniadis:1987dx}
I.~Antoniadis, J.~R.~Ellis, J.~S.~Hagelin and D.~V.~Nanopoulos,
Phys. Lett. B \textbf{194} (1987), 231-235


\bibitem{Rehman:2018nsn}
M.~U.~Rehman, Q.~Shafi and U.~Zubair,
Phys. Rev. D \textbf{97}, no.12, 123522 (2018)
doi:10.1103/PhysRevD.97.123522
[arXiv:1804.02493 [hep-ph]].

\bibitem{Pati:1974yy}
J.~C.~Pati and A.~Salam,
Phys. Rev. D \textbf{10} (1974), 275-289
[erratum: Phys. Rev. D \textbf{11} (1975), 703-703]

\bibitem{Mohapatra:1974gc}
R.~N.~Mohapatra and J.~C.~Pati,
Phys. Rev. D \textbf{11} (1975), 2558

\bibitem{Ahmed:2018jlv}
W.~Ahmed and A.~Karozas,
Phys. Rev. D \textbf{98}, no.2, 023538 (2018)
[arXiv:1804.04822 [hep-ph]].


\bibitem{Ahmed:2022vlc}
W.~Ahmed, A.~Karozas, G.~K.~Leontaris and U.~Zubair,
[arXiv:2201.12789 [hep-ph]].

\bibitem{Rehman:2012gd}
M.~U.~Rehman and Q.~Shafi,
Phys. Rev. D \textbf{86}, 027301 (2012)
[arXiv:1202.0011 [hep-ph]].

\bibitem{Rehman:2014rpa}
M.~U.~Rehman and U.~Zubair,
Phys. Rev. D \textbf{91}, 103523 (2015)
[arXiv:1412.7619 [hep-ph]].

\bibitem{Khalil:2010cp}
S.~Khalil, M.~U.~Rehman, Q.~Shafi and E.~A.~Zaakouk,
Phys. Rev. D \textbf{83}, 063522 (2011)
[arXiv:1010.3657 [hep-ph]].




\bibitem{urRehman:2006hu}
M.~ur Rehman, V.~N.~Senoguz and Q.~Shafi,
Phys. Rev. D \textbf{75}, 043522 (2007)
[arXiv:hep-ph/0612023 [hep-ph]].

\bibitem{Ahmed:2022wed}
W.~Ahmed, M.~Moosa, S.~Munir and U.~Zubair,
[arXiv:2208.11888 [hep-ph]].


\bibitem{Kibble:1982ae}
T.~W.~B.~Kibble, G.~Lazarides and Q.~Shafi,
Phys. Lett. B \textbf{113} (1982), 237-239


\bibitem{apal:2019}
A.~Pal and Q.~Shafi,
Phys.\ Rev.\  D {\bf 100}, 043526 (2019)
[arXiv:1903.05703 [hep-ph]].


\bibitem{Fukugita:1986hr}
M.~Fukugita and T.~Yanagida,
Phys. Lett. B \textbf{174}, 45-47 (1986)

\bibitem{Lazarides:1990huy}
G.~Lazarides and Q.~Shafi,
Phys. Lett. B \textbf{258}, 305-309 (1991)





\bibitem{Hill:1982iq}
C.~T.~Hill,
Nucl. Phys. B \textbf{224}, 469-490 (1983)


\bibitem{Vilenkin:1984ib}
A.~Vilenkin,
Phys. Rept. \textbf{121}, 263-315 (1985)

\bibitem{Vilenkin:2000jqa}
A.~Vilenkin and E.~P.~S.~Shellard,Cosmic Strings and Other Topological Defects, Cambridge
University Press, 2000.


\bibitem{Planck:2018jri}
Y.~Akrami \textit{et al.} [Planck],
Astron. Astrophys. \textbf{641}, A10 (2020)
[arXiv:1807.06211 [astro-ph.CO]].


















\bibitem{rehman}
M.U.~Rehman, Q.~Shafi, and J.R.~Wickman, Phys. Lett. 
B {\bf 683}, 191 (2010);
C.~Pallis and Q.~Shafi, Phys. Lett. B {\bf 725}, 327 
(2013);
W.~Buchm\"{u}ller, V.~Domcke, K.~Kamada, and K.~Schmitz, 
J. Cosmol. Astropart. Phys. {\bf 07}, 054 (2014).

\bibitem{gravitywaves}
M.U.~Rehman, Q.~Shafi, and J.R.~Wickman, Phys. Rev. D 
{\bf 83}, 067304 (2011); 
M.~Civiletti, C.~Pallis, and Q.~Shafi, Phys. Lett. B 
{\bf 733}, 276 (2014).


\bibitem{Ahmed:2022rwy}
W.~Ahmed, M.~Junaid, S.~Nasri and U.~Zubair,
Phys. Rev. D \textbf{105}, no.11, 115008 (2022)
[arXiv:2202.06216 [hep-ph]].

\bibitem{Afzal:2022vjx}
A.~Afzal, W.~Ahmed, M.~U.~Rehman and Q.~Shafi,
Phys. Rev. D \textbf{105}, no.10, 103539 (2022)
[arXiv:2202.07386 [hep-ph]].

\bibitem{Ahmed:2021dvo}
W.~Ahmed, A.~Karozas and G.~K.~Leontaris,
Phys. Rev. D \textbf{104}, no.5, 055025 (2021)
[arXiv:2104.04328 [hep-ph]].
\bibitem{Ahmed:2022rwy}
W.~Ahmed, M.~Junaid, S.~Nasri and U.~Zubair,
Phys. Rev. D \textbf{105}, no.11, 115008 (2022)
[arXiv:2202.06216 [hep-ph]].

\bibitem{bastero}
M.~Bastero-Gil, S.F.~King, and Q.~Shafi, Phys. Lett.
B {\bf 651}, 345 (2007).




\bibitem{Ade:2015lrj}
P.~A.~R.~Ade {\it et al.} [Planck Collaboration],
Astron.\ Astrophys.\  {\bf 594}, A20 (2016)
[arXiv:1502.02114 [astro-ph.CO]].

 
 \bibitem{Andre:2013afa}
 P.~Andre \textit{et al.} [PRISM],
 [arXiv:1306.2259 [astro-ph.CO]].
 
 
 \bibitem{Matsumura:2013aja}
 T.~Matsumura \textit{et al.} [Mission design of LiteBIRD],
 J. Low Temp. Phys. \textbf{176}, 733 (2014)
 [arXiv:1311.2847 [astro-ph.IM]].
 
  
  \bibitem{Finelli:2016cyd}
  F.~Finelli \textit{et al.} [CORE],
  JCAP \textbf{04}, 016 (2018)
  [arXiv:1612.08270 [astro-ph.CO]].
  
  \bibitem{Kogut:2011xw}
  A.~Kogut, D.~J.~Fixsen, D.~T.~Chuss, J.~Dotson, E.~Dwek, M.~Halpern, G.~F.~Hinshaw, S.~M.~Meyer, S.~H.~Moseley and M.~D.~Seiffert, \textit{et al.}
  JCAP \textbf{07} (2011), 025
  [arXiv:1105.2044 [astro-ph.CO]].
  
  
 \bibitem{Abazajian:2019eic}
 K.~Abazajian, G.~Addison, P.~Adshead, Z.~Ahmed, S.~W.~Allen, D.~Alonso, M.~Alvarez, A.~Anderson, K.~S.~Arnold and C.~Baccigalupi, \textit{et al.}
 [arXiv:1907.04473 [astro-ph.IM]].
 
 \bibitem{Sehgal:2019ewc}
 N.~Sehgal, S.~Aiola, Y.~Akrami, K.~Basu, M.~Boylan-Kolchin, S.~Bryan, S.~Clesse, F.~Y.~Cyr-Racine, L.~Di Mascolo and S.~Dicker, \textit{et al.}
 [arXiv:1906.10134 [astro-ph.CO]].
 
 \bibitem{SimonsObservatory:2018koc}
 P.~Ade \textit{et al.} [Simons Observatory],
 JCAP \textbf{02}, 056 (2019)
 [arXiv:1808.07445 [astro-ph.CO]].



\bibitem{Ade:2013xla}
P.~A.~R.~Ade \textit{et al.} [Planck],
``Planck 2013 results. XXV. Searches for cosmic strings and other topological defects,''
Astron. Astrophys. \textbf{571}, A25 (2014)
[arXiv:1303.5085 [astro-ph.CO]].

\bibitem{Ade:2015xua}
P.~A.~R.~Ade \textit{et al.} [Planck],
``Planck 2015 results. XIII. Cosmological parameters,''
Astron. Astrophys. \textbf{594}, A13 (2016)
[arXiv:1502.01589 [astro-ph.CO]].





\bibitem{Super-Kamiokande:2016exg}
K.~Abe \textit{et al.} [Super-Kamiokande],
Phys. Rev. D \textbf{95} (2017) no.1, 012004
[arXiv:1610.03597 [hep-ex]].

\bibitem{Nilles:1983ge}
H.~P.~Nilles,
Phys. Rept. \textbf{110} (1984), 1-162

\bibitem{Barr:2005xya}
S.~M.~Barr, B.~Kyae and Q.~Shafi,
[arXiv:hep-ph/0511097 [hep-ph]].

\bibitem{Fallbacher:2011xg}
M.~Fallbacher, M.~Ratz and P.~K.~S.~Vaudrevange,
Phys. Lett. B \textbf{705}, 503-506 (2011)


  
  \bibitem{Masoud:2019gxx}
  M.~A.~Masoud, M.~U.~Rehman and Q.~Shafi,
  JCAP \textbf{04}, 041 (2020)
  [arXiv:1910.07554 [hep-ph]].

\bibitem{Nagata:2013ive}
N.~Nagata,
``Proton Decay in High-scale Supersymmetry,''
doi:10.15083/00006623


1 	


\bibitem{Garcia-Bellido:1996egv}
J.~Garcia-Bellido and A.~D.~Linde,
Phys. Lett. B \textbf{398}, 18-22 (1997)
[arXiv:astro-ph/9612141 [astro-ph]].


\bibitem{Nakayama:2010xf}
K.~Nakayama, F.~Takahashi and T.~T.~Yanagida,
JCAP \textbf{12}, 010 (2010)
[arXiv:1007.5152 [hep-ph]].

\bibitem{Liddle:1993fq}
A.~R.~Liddle and D.~H.~Lyth,
Phys. Rept. \textbf{231}, 1-105 (1993)
[arXiv:astro-ph/9303019 [astro-ph]].
















\bibitem{Pilaftsis:1991ug} 
A.~Pilaftsis,
``Radiatively induced neutrino masses and large Higgs neutrino couplings in the standard model with Majorana fields,''
Z.\ Phys.\ C {\bf 55}, 275 (1992)
[hep-ph/9901206].

\bibitem{Tommasini:1995ii} 
D.~Tommasini, G.~Barenboim, J.~Bernabeu and C.~Jarlskog,
``Nondecoupling of heavy neutrinos and lepton flavor violation,''
Nucl.\ Phys.\ B {\bf 444}, 451 (1995)
[hep-ph/9503228].

\bibitem{Gluza:2002vs} 
J.~Gluza,
``On teraelectronvolt Majorana neutrinos,''
Acta Phys.\ Polon.\ B {\bf 33}, 1735 (2002)
[hep-ph/0201002].

\bibitem{Xing:2009in} 
Z.~z.~Xing,
``Naturalness and Testability of TeV Seesaw Mechanisms,''
Prog.\ Theor.\ Phys.\ Suppl.\  {\bf 180}, 112 (2009)
[arXiv:0905.3903 [hep-ph]].

\bibitem{Gavela:2009cd} 
M.~B.~Gavela, T.~Hambye, D.~Hernandez and P.~Hernandez,
``Minimal Flavour Seesaw Models,''
JHEP {\bf 0909}, 038 (2009)
[arXiv:0906.1461 [hep-ph]].

\bibitem{He:2009ua} 
X.~G.~He, S.~Oh, J.~Tandean and C.~C.~Wen,
``Large Mixing of Light and Heavy Neutrinos in Seesaw Models and the LHC,''
Phys.\ Rev.\ D {\bf 80}, 073012 (2009)
[arXiv:0907.1607 [hep-ph]].

\bibitem{Adhikari:2010yt} 
R.~Adhikari and A.~Raychaudhuri,
``Light neutrinos from massless texture and below TeV seesaw scale,''
Phys.\ Rev.\ D {\bf 84}, 033002 (2011)
[arXiv:1004.5111 [hep-ph]].

\bibitem{Deppisch:2010fr} 
F.~F.~Deppisch and A.~Pilaftsis,
``Lepton Flavour Violation and theta(13) in Minimal Resonant Leptogenesis,''
Phys.\ Rev.\ D {\bf 83}, 076007 (2011)
[arXiv:1012.1834 [hep-ph]].

\bibitem{Mitra:2011qr} 
M.~Mitra, G.~Senjanovic and F.~Vissani,
``Neutrinoless Double Beta Decay and Heavy Sterile Neutrinos,''
Nucl.\ Phys.\ B {\bf 856}, 26 (2012)
[arXiv:1108.0004 [hep-ph]].

\bibitem{Dev:2013oxa} 
C.~H.~Lee, P.~S.~B.~Dev and R.~N.~Mohapatra,
``Natural TeV-scale left-right seesaw mechanism for neutrinos and experimental tests,''
Phys.\ Rev.\ D {\bf 88}, no. 9, 093010 (2013)
[arXiv:1309.0774 [hep-ph]].

\bibitem{Chattopadhyay:2017zvs} 
P.~Chattopadhyay and K.~M.~Patel,
``Discrete symmetries for electroweak natural type-I seesaw mechanism,''
Nucl.\ Phys.\ B {\bf 921}, 487 (2017)
[arXiv:1703.09541 [hep-ph]].

\bibitem{Das:2017nvm} 
A.~Das and N.~Okada,
``Bounds on heavy Majorana neutrinos in type-I seesaw and implications for collider searches,''
Phys.\ Lett.\ B {\bf 774}, 32 (2017)
[arXiv:1702.04668 [hep-ph]].

\bibitem{Datta:1993nm} 
A.~Datta, M.~Guchait and A.~Pilaftsis,
``Probing lepton number violation via majorana neutrinos at hadron supercolliders,''
Phys.\ Rev.\ D {\bf 50}, 3195 (1994)
[hep-ph/9311257].

\bibitem{Panella:2001wq} 
O.~Panella, M.~Cannoni, C.~Carimalo and Y.~N.~Srivastava,
``Signals of heavy Majorana neutrinos at hadron colliders,''
Phys.\ Rev.\ D {\bf 65}, 035005 (2002)
[hep-ph/0107308].

\bibitem{Han:2006ip} 
T.~Han and B.~Zhang,
``Signatures for Majorana neutrinos at hadron colliders,''
Phys.\ Rev.\ Lett.\  {\bf 97}, 171804 (2006)
[hep-ph/0604064].

\bibitem{delAguila:2007qnc} 
F.~del Aguila, J.~A.~Aguilar-Saavedra and R.~Pittau,
``Heavy neutrino signals at large hadron colliders,''
JHEP {\bf 0710}, 047 (2007)
[hep-ph/0703261].

\bibitem{Dev:2013wba} 
P.~S.~B.~Dev, A.~Pilaftsis and U.~k.~Yang,
``New Production Mechanism for Heavy Neutrinos at the LHC,''
Phys.\ Rev.\ Lett.\  {\bf 112}, no. 8, 081801 (2014)
[arXiv:1308.2209 [hep-ph]].

\bibitem{Alva:2014gxa} 
D.~Alva, T.~Han and R.~Ruiz,
``Heavy Majorana neutrinos from $W\gamma$ fusion at hadron colliders,''
JHEP {\bf 1502}, 072 (2015)
[arXiv:1411.7305 [hep-ph]].

\bibitem{Das:2015toa} 
A.~Das and N.~Okada,
``Improved bounds on the heavy neutrino productions at the LHC,''
Phys.\ Rev.\ D {\bf 93}, no. 3, 033003 (2016)
[arXiv:1510.04790 [hep-ph]].


\bibitem{Das:2016hof} 
A.~Das, P.~Konar and S.~Majhi,
``Production of Heavy neutrino in next-to-leading order QCD at the LHC and beyond,''
JHEP {\bf 1606}, 019 (2016)
[arXiv:1604.00608 [hep-ph]].

\bibitem{Pascoli:2018heg} 
S.~Pascoli, R.~Ruiz and C.~Weiland,
``Heavy Neutrinos with Dynamic Jet Vetoes: Multilepton Searches at $\sqrt{s} = 14,~27,$ and $100$ TeV,''
arXiv:1812.08750 [hep-ph].


\bibitem{Dev:2012zg} 
P.~S.~B.~Dev, R.~Franceschini and R.~N.~Mohapatra,
``Bounds on TeV Seesaw Models from LHC Higgs Data,''
Phys.\ Rev.\ D {\bf 86}, 093010 (2012)
[arXiv:1207.2756 [hep-ph]].

\bibitem{Cely:2012bz} 
C.~G.~Cely, A.~Ibarra, E.~Molinaro and S.~T.~Petcov,
``Higgs Decays in the Low Scale Type I See-Saw Model,''
Phys.\ Lett.\ B {\bf 718}, 957 (2013)
[arXiv:1208.3654 [hep-ph]].

\bibitem{Hessler:2014ssa} 
A.~G.~Hessler, A.~Ibarra, E.~Molinaro and S.~Vogl,
``Impact of the Higgs boson on the production of exotic particles at the LHC,''
Phys.\ Rev.\ D {\bf 91}, no. 11, 115004 (2015)
[arXiv:1408.0983 [hep-ph]].

\bibitem{Das:2017zjc} 
A.~Das, P.~S.~B.~Dev and C.~S.~Kim,
``Constraining Sterile Neutrinos from Precision Higgs Data,''
Phys.\ Rev.\ D {\bf 95}, no. 11, 115013 (2017)
[arXiv:1704.00880 [hep-ph]].

\bibitem{Das:2017rsu} 
A.~Das, Y.~Gao and T.~Kamon,
``Heavy neutrino search via semileptonic Higgs decay at the LHC,''
Eur.\ Phys.\ J.\ C {\bf 79}, no. 5, 424 (2019)
[arXiv:1704.00881 [hep-ph]].



\bibitem{Cacciapaglia:2006pk}
G.~Cacciapaglia, C.~Csaki, G.~Marandella and A.~Strumia,
Phys. Rev. D \textbf{74}, 033011 (2006)
doi:10.1103/PhysRevD.74.033011
[arXiv:hep-ph/0604111 [hep-ph]].

\bibitem{Accomando:2013sfa}
E.~Accomando, D.~Becciolini, A.~Belyaev, S.~Moretti and C.~Shepherd-Themistocleous,
JHEP {\bf 1310}, 153 (2013)
doi:10.1007/JHEP10(2013)153
[arXiv:1304.6700 [hep-ph]];
W.~Abdallah, J.~Fiaschi, S.~Khalil and S.~Moretti,
JHEP {\bf 1602}, 157 (2016)
doi:10.1007/JHEP02(2016)157
[arXiv:1510.06475 [hep-ph]];
E.~Accomando, A.~Belyaev, J.~Fiaschi, K.~Mimasu, S.~Moretti and C.~Shepherd-Themistocleous,
JHEP {\bf 1601}, 127 (2016)
doi:10.1007/JHEP01(2016)127
[arXiv:1503.02672 [hep-ph]].
J.~Y.~Araz, G.~Corcella, M.~Frank and B.~Fuks,
JHEP {\bf 1802}, 092 (2018)
[arXiv:1711.06302 [hep-ph]].
\bibitem{ATLAS:2017wce} 
The ATLAS collaboration [ATLAS Collaboration],
ATLAS-CONF-2017-027.

	
	
	
\end{thebibliography}
\end{document}